\pdfoutput=1
\documentclass[twocolumn]{aastex62}

\usepackage{amsmath,graphicx,multirow}
\hypersetup{breaklinks,colorlinks,allcolors=blue}
 
\received{2018 September 11}
\revised{2018 September 28}
\accepted{2018 September 30}
\submitjournal{ApJ}

\shorttitle{Stellar Population Diagnostics of the Massive Star Binary Fraction}
\shortauthors{Dorn-Wallenstein \& Levesque}

\begin{document}

\title{Stellar Population Diagnostics of the Massive Star Binary Fraction}

\correspondingauthor{Trevor Z. Dorn-Wallenstein}
\email{tzdw@uw.edu}

\author[0000-0003-3601-3180]{Trevor Z. Dorn-Wallenstein}
\affiliation{University of Washington Astronomy Department \\
Physics and Astronomy Building, 3910 15th Ave NE  \\
Seattle, WA 98105, USA} 

\author[0000-0003-2184-1581]{Emily M. Levesque}
\affiliation{University of Washington Astronomy Department \\
Physics and Astronomy Building, 3910 15th Ave NE  \\
Seattle, WA 98105, USA}

\begin{abstract}

Populations of massive stars are directly reflective of the physics of stellar evolution. Counting subtypes of massive stars and ratios of massive stars in different evolutionary states have been used ubiquitously as diagnostics of age and metallicity effects. While the binary fraction of massive stars is significant, inferences are often based upon models incorporating only single-star evolution. In this work, we utilize custom synthetic stellar populations from the Binary Population and Stellar Synthesis (BPASS) code to determine the effect of stellar binaries on number count ratios of different evolutionary stages in both young massive clusters and galaxies with massive stellar populations. We find that many ratios are degenerate in metallicity, age, and/or binary fraction. We develop diagnostic plots using these stellar count ratios to help break this degeneracy, and use these plots to compare our predictions to observed data in the Milky Way and the Local Group. These data suggest a possible correlation between the massive star binary fraction and metallicity. We also examine the robustness of our predictions in samples with varying levels of completeness. We find including binaries and imposing a completeness limit can both introduce $\gtrsim0.1$ dex changes in inferred ages. Our results highlight the impact that binary evolution channels can have on the massive star population.

\end{abstract}


\section{Introduction} \label{sec:intro}

Comparing theoretical and observed populations of massive stars can be an incredibly powerful tool for understanding stellar evolution. Massive stars are luminous, and can be easily seen in the Local Group; photometric catalogs are readily available \citep[e.g., the Local Group Galaxy Survey, LGGS; ][]{massey06,massey07}, from which massive stars can be selected after filtering for foreground contaminants \citep[e.g., ][]{massey09}. The relative abundance of various subtypes of massive stars can then be used as a probe of stellar physics. Reproducing the observed data reflects our understanding of the relative lifetimes of these evolutionary phases, and therefore our ability to predict the impact of massive stars on their surroundings --- e.g., chemical yields, ionizing radiation, and mechanical feedback. However, most stellar evolution models assume that stars evolve in relative isolation, without any influence from a binary companion.

In a radial velocity survey of 71 Galactic O stars in nearby open clusters, \citet{sana12} directly searched for binary systems, and used their results to infer the intrinsic binary fraction $f_{bin}$ as well the distributions of binary parameters. They reported a tendency for binaries to favor close systems with mass ratios drawn from a uniform distribution. They also found a high binary fraction $f_{bin}>70\%$ when including longer-period systems, and subsequent work \citep[see e.g.,][]{duchene13,sana14,moe17} has also found evidence that the evolution of massive stars is dominated by binary interactions in close systems. However, other observations place the short-period binary fraction for O-type stars at $\sim$30-35\% \citep[e.g., ][]{garmany80,sana13}. For post-main-sequence massive stars, the observed binary fraction for Wolf-Rayet stars is $\sim$30\% \citep{neugent14}, while the binary fraction of yellow and red supergiants is still unknown \citep{levesque17}. Furthermore, while observations are biased towards finding more massive companions, secondaries with low mass ratios are still capable of significantly altering the evolution of their primary at short orbital periods \citep{eldridge17}. 

Massive stars in interacting binary systems face drastically differing evolutionary pathways than their single-star cousins, causing ensembles of binary stars to appear notably different as a function of age, metallicity, and the underlying statistical distributions of the binary parameters of the systems. Measuring these properties directly is possible; however, such surveys are time intensive, and require detailed understandings of the completeness of the survey and the sensitivity of the observational method to systems of varying periods, inclinations, and mass ratios. Correcting for these effects in small samples can be difficult, making it hard to generalize the results to the entire population of massive stars. Thus any inferences made about young stellar populations are inherently polluted by unresolved binaries that have not been accounted for \citep{demink14}.

The predicted number of almost every subtype of massive star depends upon binarity. Perhaps the most notable effect is an increase in the expected number of stripped-envelope stars (i.e., Wolf-Rayet stars, WRs) with $f_{bin}$, which occurs due to Roche-Lobe Overflow (RLOF) onto the secondary star. Other, more subtle, effects can alter the number of massive stellar subtypes observed through time, e.g., red supergiants (RSGs), yellow supergiants (YSGs), blue supergiants (BSG) and the various WR subtypes (WC, WN, etc.). Indeed, recent work has argued that entire subclasses of massive stars may exclusively be the product of binary evolution (e.g., Luminous Blue Variables, LBVs; see \citealt{smith15,humphreys16}). One may conclude then that using stellar count diagnostics as a probe of stellar physics is hopeless in the presence of binary stars with unknown properties. More optimistically, we seek to understand the effect of binaries on star count diagnostics to determine if they can be used to disentangle binary effects from single-star evolution.

This paper is laid out as follows. In \S\ref{sec:method}, we give a brief overview of the effect of binary interactions, describing the Binary Population and Spectral Synthesis (BPASS) code and the population synthesis method we employ to generate theoretical predictions for the abundance of various subtypes. In \S\ref{sec:ratios} we describe various ratios of these subtypes that are sensitive to age, metallicity, and binary fraction. \S\ref{sec:data} describes how these ratios can be applied to real data, while \S\ref{sec:luminosity} includes our presciption for handling incomplete samples of massive stars, and the effect of incompleteness on the inferred results. We apply these ratios to populations with complicated star-formation histories in \S\ref{sec:complex} before concluding in \S\ref{sec:conclusion}.

\section{Creating Theoretical Populations With a Physical Treatment of Binaries}\label{sec:method}

\subsection{Binary Evolution and BPASS}

Stars born in close binaries interact primarily via tides and mass-transfer \citep{hurley02}, the latter of which can occur via both stellar winds and RLOF. Both mechanisms can change the angular momentum of the system, affecting the orbital separation of the system and the rotation speeds of the individual stars \citep{demink13}, which in turn can lead to further interactions, including those where the stars come into contact with each other. In the most extreme cases, the system enters a brief common envelope phase, which may be followed by a merger, depending on the orbital energy of the system \citep{paczynski76,demink14}. The effects of these interactions on the evolution of both stars in the system is heavily dependent on the evolutionary state of each star \citep{langer12}, which depends on the initial period and mass ratio of the system. This leaves a large parameter space that must be fully explored in order to sample the entire range of binary effects. 

\begin{figure}[ht!]
\plotone{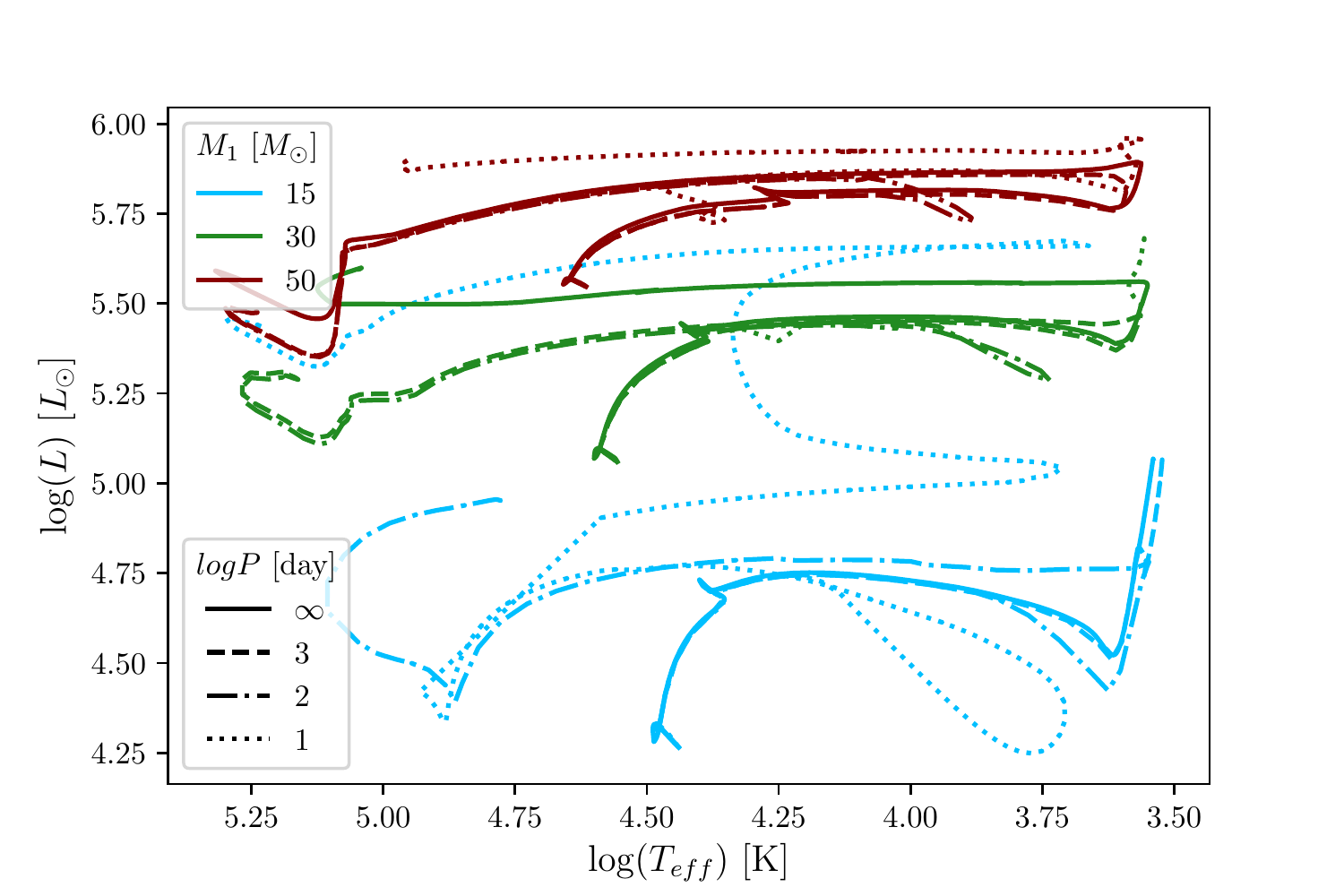}
\caption{BPASS solar metallicity stellar evolution tracks for 15 (blue), 30 (green), and 50 (red) $M_\odot$ primary stars, assuming a mass ratio $q=0.9$, and initial orbital periods of 1000 (dashed), 100 (dash-dotted), and 10 days (dotted) respectively. The corresponding single-star evolution track is shown with the solid line.}\label{fig:hr_diag}
\end{figure}

The Binary Population and Spectral Synthesis code \citep[BPASS, ][]{eldridge17,stanway18} incorporates many of these effects in a custom stellar evolution code that is evaluated for single and binary stars on a dense grid\footnote{Details on the grids of parameter values and more can be found in the BPASS v2.2 User Manual, currently hosted online at \url{https://drive.google.com/file/d/1IYCYf5Bxt1WmqPuFTYLQ7kpN-hKY2SAp/view?usp=sharing}} of initial primary and secondary masses ($M_1$ and $M_2$), initial periods $P$, and mass ratios ($q\equiv M_2/M1$) at 12 metallicities. We express the metallicity as a mass fraction $Z$, and BPASS adopts metallicities in the range $10^{-5}\leq Z \leq 0.04$. Note that for the duration of this paper, we assume solar metallicity $Z_\odot = 0.014$ \citep{asplund09}. Figure \ref{fig:hr_diag} shows solar metallicity evolutionary tracks from BPASS v2.2 for primary stars of initial mass $M_1$ between 15 and 50 $M_\odot$, companions with mass ratio $q = 0.9$, and initial orbital periods between 10 and 1000 days, as well as the corresponding single-star evolution track ($P\rightarrow\infty$). At the widest orbital separations, the primary stars evolve more or less identically to their single counterparts until the very end of their lives, where they fill their Roche lobes as yellow or red supergiants. This is most noticeable for the 30 $M_\odot$ tracks, where the primary of the $P = 10^3$ day binary only fills its Roche Lobe when it is close to the Hayashi limit. The subsequent mass transfer reduces the luminosity of the primary (see Figure 13.1 of \citealt{lamers99}, with data from \citealt{deloore78}), and causes the primary to end its life as a lower luminosity Wolf-Rayet star.

At progressively shorter periods, the effects of mass transfer become increasingly extreme. This is especially drastic for the 10 $M_\odot$ models, where mass transfer begins to occur earlier and earlier in the star's post-main sequence life, drastically altering its evolution. In the most extreme case presented ($P = 10$ days), the primary experiences multiple episodes of mass transfer both to and from the secondary, ultimately becoming an incredibly luminous Wolf-Rayet star instead of reaching the Hayashi limit and ending its life as a red supergiant as would be expected for an isolated 10 $M_\odot$ star. We note that these are extreme examples chosen to illustrate the range of behavior that occurs in close binaries for specific combinations of system parameters; each individual model is ultimately assigned a very small weight in the ensuing population synthesis due to the large number of models (see \S\ref{subsec:popsyn}).

If massive stars truly favor high binary fractions and short orbital periods then very few stars will evolve completely free from the influence of a companion. This has a drastic effect on the relative numbers of stars of a given subtype in a population. For example, the cluster Westerlund 1 is a single-age ($\sim5$ Myr, \citealt{kudryavtseva12}) massive cluster known to contain both red supergiants and Wolf-Rayet stars \citep{clark05}. Single-star evolution predicts that these stages are evolved from stars in two almost entirely disjoint sets of initial masses, implying that single-aged clusters containing both RSGs and WRs should only exist for an incredibly narrow window of time after an initial starburst. However, allowing for the formation of Wolf-Rayet stars via RLOF-induced channel increases the overlap in the initial masses of RSG and WR progenitors.

\subsection{Population Synthesis}\label{subsec:popsyn}

With the complete set of single and binary stellar evolution tracks, we assembled synthetic populations by weighting each model according to the likelihood that it would be formed in an instantaneous burst of star formation. When considering only single stars, the weighting is calculated according to an initial mass function (IMF), the probability $\Phi(M)$ of a star being formed at a given mass. $\Phi$ is typically parametrized as a power law or broken power law. BPASS allows for population synthesis assuming one of nine IMFs, including IMFs with low-mass exponential cutoffs \citep{chabrier03}, and the classical Salpeter IMF with a slope of -2.35 \citep{salpeter55}. We adopt the BPASS default, which is a broken power law with slope -1.3 below 0.5 $M_\odot$, and a slope of -2.35 for higher masses, with a maximum mass of 300 $M_\odot$. Because we are mostly considering massive stars, the shape of the low-mass IMF should have little effect on our results.

When adding binary stars, individual models must also be weighted according to the distributions of the fundamental natal parameters $P$ and $q$ of the binary system. BPASS v2.2 adopts the distribution parameters from \citet{moe17}, who found that these distrubitions are interrelated with, for example, the power law slope and twin ($q=1$) fraction of the mass ratio distribution depending on the initial mass and period\footnote{The exact distribution parameters can be found in Table 13 of \citet{moe17}}. 

Finally the weightings are normalized to ensure that the entire population forms $10^6 M_\odot$ of stars. For each metallicity, we create two synthetic populations: one composed entirely of single stars using the input files provided in the BPASS v2.2 data release, and one composed entirely of binary stars using custom input files provided by the BPASS team (J. J. Eldridge 2018, private communication). While no $f_{bin}=0$ or 1 populations have been observed, creating these populations allows us to generate results with tailored intermediate values of $f_{bin}$ by mixing both populations accordingly. Note that the binary input files provided in the BPASS v2.2 data release assume the binary fractions found by \citealt{moe17}, which would enforce an implicit maximum $f_{bin}$ on our results; we instead use our custom $f_{bin}=1$ population to avoid biasing our results. 

\subsection{Number Counts vs. Time}\label{subsec:numbers}

\begin{figure}[ht!]
\plotone{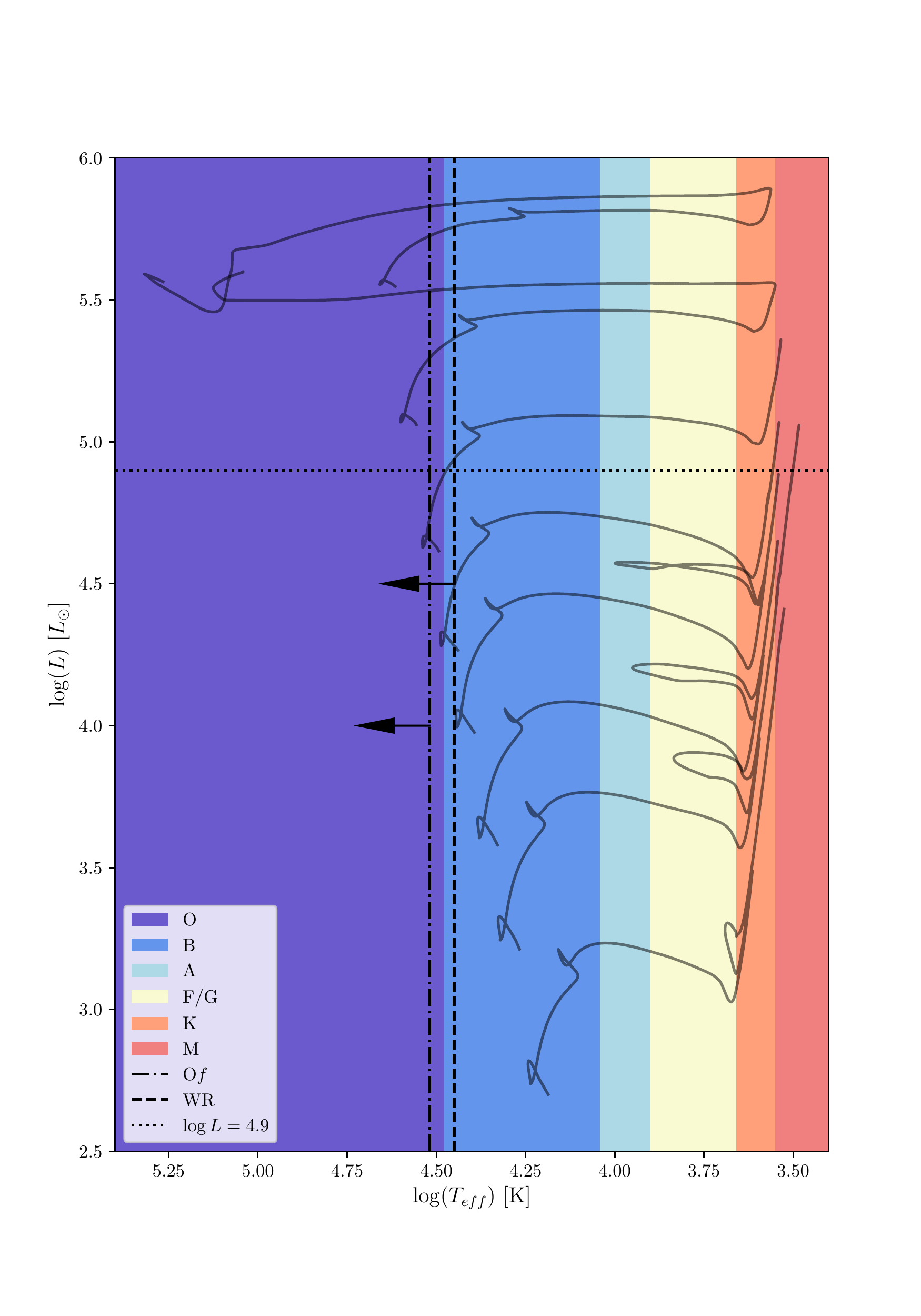}
\caption{Visualization of the criteria used to count massive stellar subtypes from \citet{eldridge17}. Spectral types are indicated by the colored patches. The minimum temperatures for O$f$ and WR stars are shown by the dash-dotted and dashed lines respectively; additional criteria, and the criteria for various WR subtypes are in Table \ref{tab:count_criteria}. For comparison, single-star solar metallicity BPASS tracks from 5 to 50 $M_\odot$ are shown in gray.}\label{fig:hr_diag_crit}
\end{figure}

We now examine each stellar evolution track to determine the evolutionary phases that it goes through by assigning a subtype to all timesteps in the model, using the model parameters listed in Table \ref{tab:mod_params}; we largely adapt the classification scheme from \citet{eldridge17}. First a check is done on $\log(T_{eff})$ and $X$ to determine if the star is a Wolf-Rayet star --- i.e., $\log(T_{eff})\geq4.45$ and $X\leq0.4$. If it is and $X > 10^{-3}$, it is a WNH star; otherwise, it is classified as a WN or WC star depending on the ratio of $C+O$ to $Y$. If a star is not a WR, a $\log(T_{eff})$ and $\log(g)$ check is performed to determine if the star is a O$f$ star --- i.e., an evolved O star with a particularly strong stellar wind. These rare stars are included as a separate class because they are particularly strong sources of \ion{He}{2} emission lines \citep{brinchmann08}. If the star is neither a WR nor a O$f$ star, it is then assigned a classical MK spectral type based on its effective temperature. The exact numerical criteria used for our classification are listed in Table \ref{tab:count_criteria}. Figure \ref{fig:hr_diag_crit} serves as an illustration of the various temperature criteria used, compared to evolutionary tracks for single stars with initial masses between 5 and 50 $M_\odot$. 

\begin{deluxetable}{lll}
\tabletypesize{\normalsize}
\tablecaption{BPASS model parameters used to label timesteps with an evolutionary phase\label{tab:mod_params}.}
\tablehead{\colhead{Parameter} &
\colhead{Description} & \colhead{Unit}} 
\startdata
$\log(L)$ & Logarithm of the luminosity & $L_\odot$ \\
$\log(T_{eff})$ & Logarithm of the effective temperature & K \\
$\log(g)$ & Logarithm of the surface gravity & cm s$^{-2}$ \\
$X$ & Hydrogen Surface Mass Fraction & - \\
$Y$ & Helium Surface Mass Fraction & - \\
$C$ & Carbon Surface Mass Fraction & - \\
$O$ & Oxygen Surface Mass Fraction & - \\
\enddata
\end{deluxetable}

Once a timestep is assigned a label, it is then assigned to at least one of 51 time bins that are logarithmically-spaced between $10^6$ and $10^{11}$ years in 0.1 dex increments. The weight of that model from the input file is then adjusted by the size of the model timestep (accounting for the fact that some timesteps cross the boundaries of the logarithmic time bin, and thus contribute to two time bins in differing amounts), and its final weight added to an entry in an array corresponding to its assigned label, time bin, and luminosity. We then create output arrays for each subtype by summing the array over the luminosity axis. We also create outputs assuming minimum luminosities for each subtype in 0.1 dex steps between $\log(L) = 3$ and 6. All of the arrays have been compiled into a single file, which we make available online.

Finally, as a post-processing step for this paper, we label stars above $\log(L)=4.9$ as supergiants, classifying O, O$f$, B, and A stars as BSGs, F and G stars as YSGs, and K and M stars as RSGs. As discussed by \citet{eldridge17}, this is based on the luminosity criterion used by \citet{massey03} to ensure that lower-mass AGB stars were not included in their sample of RSGs, and applied to the rest of the supergiants for consistency. 

For most of our analyses, we apply the same luminosity cutoff when considering the number of Wolf-Rayet stars in a population. As discussed previously, binary interactions are capable of stripping low mass stars that would be otherwise incapable of losing that much mass through stellar winds or instabilities alone. This results in a large number of ``Wolf-Rayet'' stars at ages well older than when the last WR stars are expected to disappear. These stars may appear as both binary systems and single stars (in the case of stripped secondaries), and exhibit a range of spectra from classical WR spectra to hot subdwarfs \citep{gotberg18}. Very few such systems have been found --- e.g. $\phi$ Persei \citep{gies98}, FY CMa \citep{peters08}, 59 Cyg \citep{peters13}, 60 Cyg \citep{wang17}, HD 45166 \citep{steiner05,groh08}. This may be due to detectability issues (as the ``Wolf-Rayet'' primary can be far less luminous than the mass-gaining secondary), or a lack of atmospheric models for these stars. The luminosity cutoff for WRs attempts to mitigate this issue; however, BPASS still predicts the existence of luminous yet low-mass WRs well after ages of 10 Myr. No WRs have been found in intermediate-age clusters, which is in tension with a high binary fraction for massive stars.

\begin{deluxetable}{ll}
\tabletypesize{\normalsize}
\tablecaption{Criteria used to label regions of the HR diagram to classify evolution tracks. Adapted from Table 3 of \citet{eldridge17}\label{tab:count_criteria}. The luminosity cutoff for WR stars is not always applied; see \S\ref{sec:luminosity} for details.}
\tablehead{\colhead{Label} &
\colhead{Criteria}} 
\startdata
\multirow{2}{*}{WNH} & \multirow{2}{*}{\parbox{4cm}{$\log(T_{eff}) \geq 4.45$ \\ $X \leq 0.4$}} \\
\\
\multirow{3}{*}{WN} & \multirow{3}{*}{\parbox{5cm}{$\log(T_{eff}) \geq 4.45$ \\ $X \leq 10^{-3}$ \\ $(C + O))/Y \leq 0.03$}} \\
\\
\\
\multirow{3}{*}{WC} & \multirow{3}{*}{\parbox{5cm}{$\log(T_{eff}) \geq 4.45$ \\ $X \leq 10^{-3}$ \\ $(C + O))/Y > 0.03$}} \\
\\
\\
O & $\log(T_{eff}) \geq 4.48$  \\
\multirow{2}{*}{O$f$} & \multirow{2}{*}{\parbox{5cm}{$\log(T_{eff}) \geq 4.519$ \\ $\log(g) > 3.676 \log(T_{eff}) + 13.253$}} \\
\\
B & $4.041 \leq \log(T_{eff}) < 4.48$ \\
A & $3.9 \leq \log(T_{eff}) < 4.041$ \\
F/G & $3.66 \leq \log(T_{eff}) < 3.9$ \\
K & $3.55 \leq \log(T_{eff}) < 3.66 $ \\
M & $\log(T_{eff}) < 3.55$ \\
\multirow{2}{*}{BSG} & \multirow{2}{*}{\parbox{4cm}{O + O$f$ + B + A \\ $\log(L) \geq 4.9$}} \\
\\
\multirow{2}{*}{YSG} & \multirow{2}{*}{\parbox{4cm}{F/G \\ $\log(L) \geq 4.9$}} \\
\\
\multirow{2}{*}{RSG} & \multirow{2}{*}{\parbox{4cm}{K + M \\ $\log(L) \geq 4.9$}} \\
\\
\multirow{2}{*}{WR} & \multirow{2}{*}{\parbox{4cm}{WNH + WN + WC \\ $\log(L) \geq 4.9$}} \\
\\
\enddata
\end{deluxetable}

\begin{figure*}[p!]
\centering
\includegraphics[scale=0.6,angle=90]{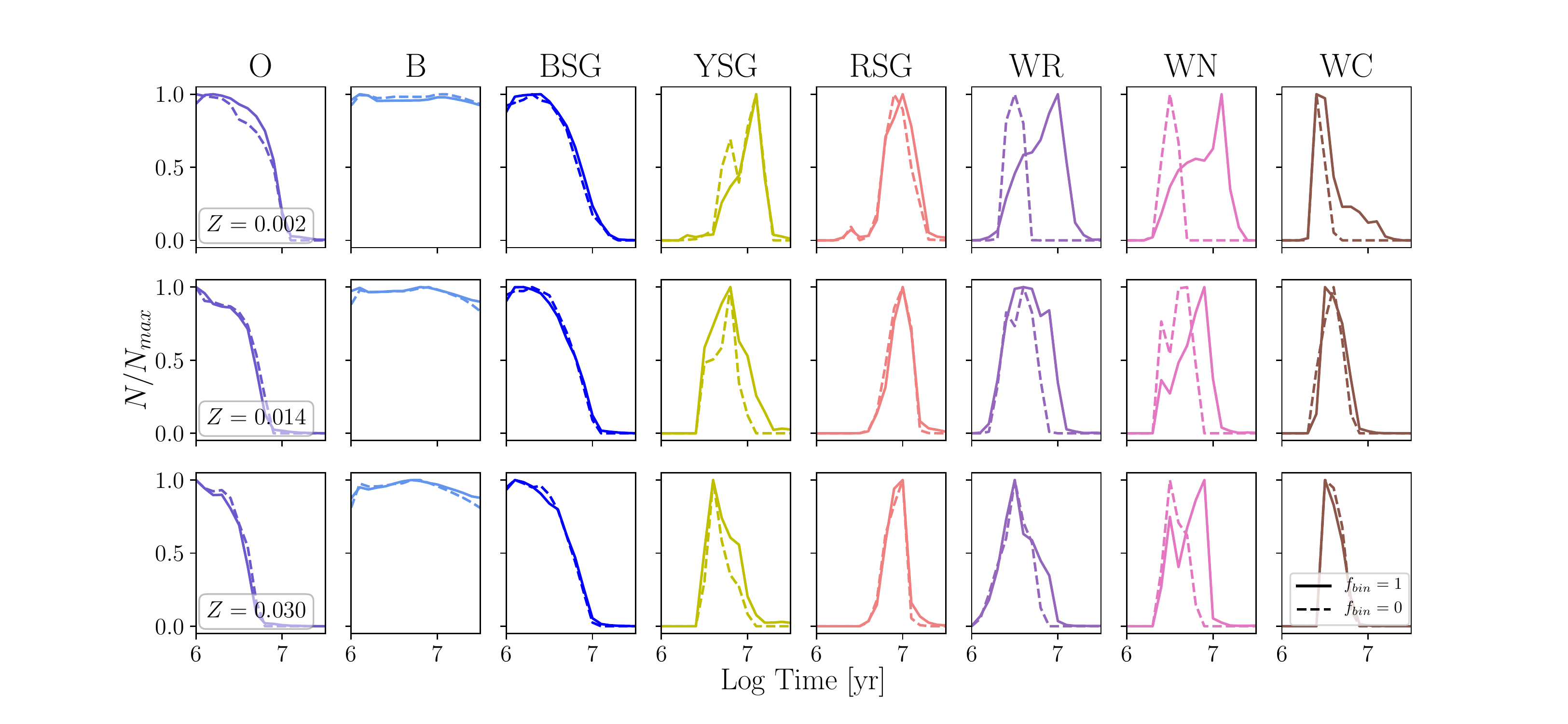}
\caption{Number of various stellar subtypes in the $Z=0.002$ (top), $Z=0.014$ (middle), and $Z=0.03$ (bottom) populations between $10^6$ and $10^{7.5}$ years, scaled so that the maximum number of any subtype is 1. The solid and dashed lines indicate the $f_{bin}=1$ and $f_{bin}=0$ populations respectively.}\label{fig:counts_all_met}
\end{figure*}

Figure \ref{fig:counts_all_met} shows the predicted number counts for times between $10^6$ and $10^{7.5}$ years at three example metallicities for subtypes O, B, BSG, YSG, RSG, WR, WN, and WC. We then find $N_{max}$, the maximum expected number of stars of each subtype at both values of $f_{bin}$, and scale by the appropriate $N_{max}$ so the maximum number of each subtype at each $f_{bin}$ is 1 for clarity. The $f_{bin} = 1$ and $f_{bin} = 0$ populations are indicated with solid and dashed lines respectively. While neither is representative of a physical sample of stars, the comparison between the two is useful for understanding the effects of binarity on a population. 

\begin{figure}[hb!]
\plotone{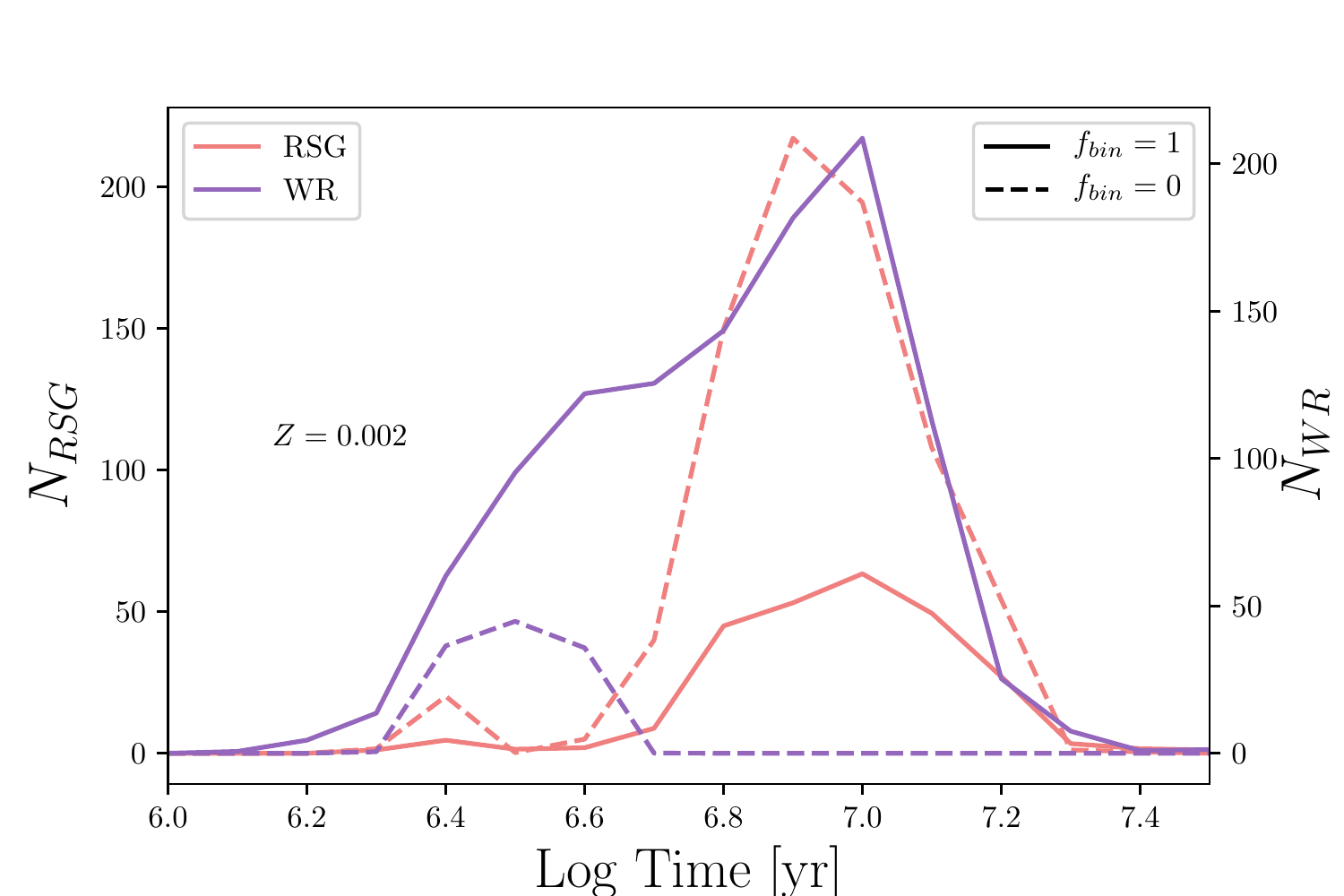}
\caption{Absolute number of RSGs and WRs per $10^6$ $M_\odot$ of stellar mass created for both binaries (solid) and singles (dashed) at $Z=0.002$}\label{fig:rsgwr_counts_002}
\end{figure}

As expected, the Wolf-Rayet stages are most heavily affected: WRs at $f_{bin}=1$ appear slightly earlier, while the age at which the most WRs are predicted is significantly later. There is also a converse effect on the RSGs, which is most noticeable at low metallicity, because the stellar-wind channel of WR creation is diminished, allowing the the effects of binarity on the WR population to dominate. Because of our $\log(L)\geq4.9$ cutoff, the WRs created by binary interactions are massive ($\geq8$ $M_\odot$), and would thus turn into RSGs if they were single stars. In Figure \ref{fig:rsgwr_counts_002}, we show the total number of RSGs (red) and WRs (purple) per $10^6$ $M_\odot$ of stellar material created in both the single (dashed) and binary (solid) populations at $Z=0.002$. In the $f_{bin}=0$ population, we see the expected behavior: there are far fewer WRs created, and they coexist with RSGs for a very narrow window of time. However, in the $f_{bin}=1$ population, the number of RSGs is suppressed by a factor of $\sim4$, and we see WRs at far later times.

We note that we do not include Luminous Blue Variables (LBVs) in our classification scheme. Due to their eruptive outbursts, LBVs are certainly important to stellar populations. However, the term has been applied to a set of objects with a wide variety of photometric and spectroscopic behavior \citep{conti84}, such that the exact definition of what is and isn't an LBV often varies from source to source, and only tens of {\it confirmed} LBVs (i.e., those which have been observed in a S Dor-type outburst) exist in the entire Local Group \citep{vangenderen01,massey07b,richardson18}. Additionally, distances to a set of LBVs and LBV candidates from the second data release of the Gaia survey \citep{gaia18} were derived by \citet{smith18}. In many cases, the updated distances are smaller than those previously reported, implying that LBVs may occupy a different region of the HR diagram than is often assumed. Due to the present uncertainty in the evolutionary status of LBVs, and the lack of a clear consensus in how to observationally classify a statistically significant number of them without long-term monitoring, we choose to not consider the LBV evolutionary phase in our analysis. 

We can now construct the expected number counts for realistic populations with a given $f_{bin}$ by mixing the two populations in proportion. However, these values are all relative to the total mass formed in a population, $M_*$. Estimating $M_*$ can be a difficult exercise, as it can depend heavily on the lower-mass IMF, as well as the measured age and distance to the population. As we demonstrate, measuring the exact age of a population can be complicated by the presence of stellar binaries. Thus instead of comparing our number count predictions directly to populations, we can construct diagnostic ratios using massive stars. These ratios are independent of both the total mass and shape of the low-mass IMF. 

\begin{figure*}[!ht]
\plotone{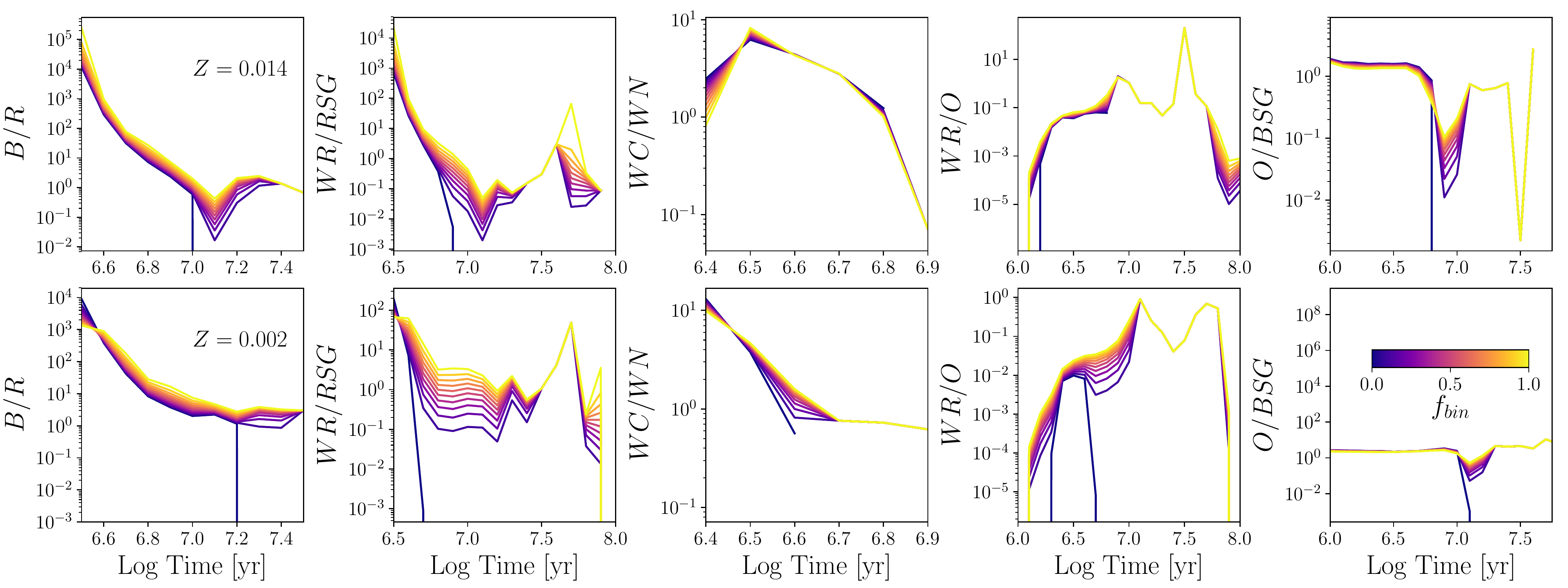}
\caption{ From left to right: $B/R$, $WR/RSG$, $WC/WN$, $WR/O$, and $O/BSG$ vs. log time at $Z=0.014$ (top row) and $Z=0.002$ bottom row, calculated for $f_{bin}$ between 0 and 1, as indicated by the colorbar on the top right. Axis limits on the abscissa are chosen to highlight the timescales on which these ratios are most dependent on varying $f_{bin}$. For all subtypes except O stars, a minimum luminosity of $\log(L)=4.9$ is enforced.}\label{fig:allratios_allmet}
\end{figure*}

\section{Diagnostic Ratios}\label{sec:ratios}

We first construct the predicted number counts for subtypes in a population with a given $f_{bin}$. We calculate the abundance of a subtype $S$ at time $t$ and metallicity $Z$ as
\begin{equation}
S(t,f_{bin},Z) = f_{bin}S_b(t,Z) + (1-f_{bin})S_s(t,Z)
\end{equation}
where $S_b$ and $S_s$ are the abundances in the $f_{bin}=1$ and $f_{bin}=0$ populations respectively. We begin by looking only at simple stellar populations (SSPs, i.e., instantaneous bursts of star formation) to determine the effect of age, metallicity, and varying $f_{bin}$, before examining more complicated populations. Figure \ref{fig:allratios_allmet} shows the values of five different ratios vs. time for SSPs with solar metallicity (top row) and $Z=0.002$ (approximately the metallicity of the SMC, bottom row), and binary fractions between 0 (purple) and 1 (yellow) as indicated by the colorbar. The bounds of the time axis have been chosen to highlight the time range during which each ratio is most dependent on $f_{bin}$.

\subsection{$B/R$}

One of the most frequently used diagnostics is the ratio of the number of blue supergiants to red supergiants ($B/R$). Its earliest uses were to corroborate the then-putative metallicity gradient in M33 \citep{walker64,vandenbergh68}. While the trend of $B/R$ increasing with increasing metallicity stymied theoretical models \citep{langer95}, it has still been used as a metallicity diagnostic by multiple studies \citep{massey03AR}. $B/R$ is mostly sensitive to the physics governing a star's rightward evolution in the HR diagram after the main sequence, and thus is dependent on rotation and convection for single stars. Note that subsequent leftward movement in the HR diagram occurs during the final stages of a star's life and is quite rapid.

Because massive stars evolve at approximately constant luminosity, and reach their coolest temperatures (i.e., largest radii) during their first crossing of the HR diagram, the first instance of RLOF for a binary must occur during this initial rightward movement. This interrups the star's normal evolution, and causes it to evolve blueward on the HR diagram. Therefore, it would make sense that binary interactions reduce the number of red supergiants, increasing $B/R$. The first column of Figure \ref{fig:allratios_allmet} shows the predicted $B/R$ values in our SSPs. We find that binarity does increase $B/R$ at most times by factors of 2-10. Considering that errors on number count ratios in star clusters can be an order of magnitude or more in all but the most massive clusters with IMFs that are well populated out to tens of $M_\odot$, measuring this effect requires exquisite statistics. However, $B/R$ varies by many orders of magnitude as a function of time, implying that it is a much better age diagnostic.

\subsection{Wolf-Rayet Ratios}

After $B/R$, perhaps the most-used number count ratios involve Wolf-Rayet stars. In the single-star paradigm (the ``Conti scenario'', \citealt{conti83}), they evolve from the most massive progenitors, and the full sequence from WN to WC/WO stars are thought to be a progression of increasingly stripped stellar envelopes. As subsequent layers are revealed, the products of more and more advanced nuclear fusion stages that have been mixed to those layers are revealed. Thus WRs are useful probes of extremely rapid mass loss. WRs have an observed binary fraction of $\sim$30\% \citep{neugent14}, to say nothing of the intrinsic binary fraction or WRs that originated as secondary stars of systems that have since been disrupted by supernovae. It is thus important to discuss WR-based diagnostic ratios in the context of stellar binaries. 

\subsubsection{$WR/RSG$}

WRs and RSGs are thought to evolve from progenitors with two mostly disjoint sets of initial masses, so their coexistance in a star cluster only occurs for an incredibly narrow window in time (e.g., the dashed lines in Figure \ref{fig:rsgwr_counts_002}). Thus, with the notable exception of Westerlund 1, which we discuss later in this section, $WR/RSG$ has most often been used in the literature as a metallicity diagnostic in galaxies: \citet{maeder80} note that $WR/RSG$ changed by factors of up to 90 in the Milky Way as a function of galactocentric distance between 7 and 13 kpc. Moreover, they proposed that $WR/RSG$ (or more accurately, its inverse) is an even more sensitive metallicity diagnostic than $B/R$. This is because the relative abundance of both subtypes is highly sensitive to the exact mass ranges of their progenitors, which in turn is affected by metallicity-dependent mass-loss.

As discussed in \S\ref{subsec:numbers}, binary interactions have an incredibly drastic effect on the relative numbers of both subtypes, especially at low metallicity. Thus it is unsurprising that the behavior of $WR/RSG$ is {\it incredibly} dependent on $f_{bin}$. The second column of Figure \ref{fig:allratios_allmet} shows $WR/RSG$ for our SSPs. As expected, at $f_{bin}=0$, $WR/RSG\rightarrow0$ by $\sim5$ Myr. However, once binaries are included, more WRs are produced, so $WR/RSG$ has defined values well after this time. Indeed, $WR/RSG$ takes on values spanning multiple orders of magnitude as a function of both age and $f_{bin}$. Issues of ``missing'' old WRs notwithstanding, we note for now that if significant numbers of these WRs produced through binary evolution channels are found in populations with ages of a few 10s of Myr, $WR/RSG$ can be a powerful diagnostic of both $f_{bin}$ and age in SSPs.

\subsubsection{$WC/WN$}

A second often-used diagnostic, $WC/WN$, uses only the relative abundance of WR subtypes. Compared to the rest of the ratios discussed, WC and WN stars arise from a mostly-overlapping set of initial masses (at least from the single-star perspective). Most interestingly, it is sensitive only to the lifetimes of WR phases, and should be mostly independent of both the IMF and which channel produces WRs. Thus, as proposed by \citet{vanbeveren80} and \citet{hellings81}, $WC/WN$ is solely a function of the metallicity and temperature dependence of Wolf-Rayet winds. The third column of Figure \ref{fig:allratios_allmet} shows $WC/WN$ vs. time at solar and subsolar metallicity for varying binary fraction. As expected, there is minimal dependence on $f_{bin}$ at almost all times, except in the lower metallicity population for a brief window around $\log{t}=6.6$. Thus, for most metallicities/ages, $WC/WN$ should indeed be a useful diagnostic, free from the influence of unresolved binaries.

\subsubsection{$WR/O$}

A third diagnostic, $WR/O$, is a probe of a large swath of the mass spectrum of massive stars. Both Galactic WR catalogs \citep[e.g., ][]{vanderhucht01} and surveys of the Local Group \citep{massey06,massey07} have made data available in environments spanning a wide range of stellar masses, metallicities, and star formation histories. However, like $WR/RSG$, $WR/O$ is especially succeptible to contamination by binaries \citep{maeder91}. The fourth column of Figure \ref{fig:allratios_allmet} shows $WR/O$ vs. time for different metallicities and binary fractions. While difficult to see due to the large y-axis scale of the plot, $WR/O$ is indeed affected by including binaries at both early and late times.

\subsection{$O/BSG$}

Finally, we introduce a ratio that is rarely discussed in the literature: $O/BSG$. This ratio is largely sensitive to the spectral type of the main sequence turnoff, and thus main sequence lifetimes. The final column of Figure \ref{fig:allratios_allmet} shows $O/BSG$ vs. time. For most ages, $O/BSG$ is insensitive to $f_{bin}$, and generally declines from early to late times as the turnoff moves to later spectral types. However, for a small window around $\log{t}\approx6.75/7$ ($Z=0.014$/$Z=0.002$ respectively), $O/BSG$ exhibits $f_{bin}$-dependent behavior. This is likely due to stars that experience moderate amounts of RLOF, and evolve blueward, but haven't lost enough to of their H envelopes to become WR stars. These stars are then classified as O, increasing $O/BSG$.

\section{Comparisons with Real Data}\label{sec:data}

Given a complete sample of massive stars in a population, it is possible to calculate stellar count ratios to compare to predictions. However, because the different subtypes require various and typically time-intensive methods for discovery and classification, it is often the case that the data for individual types of stars must be assembled from a variety of inhomogeneous sources. Thus only a few subtypes may have been cataloged, from which only a few ratios can be calculated. Therefore it is critically important to choose ratios that are best suited to the population under consideration --- i.e., suitable diagnostics of age, metallicity, or $f_{bin}$. 

Massive stars are rare, and the abundance of evolved massive stellar subtypes is subject to Poisson noise in star clusters with $M_*\sim10^5$ $M_\odot$ at most. Thus, great care must be made when comparing the theoretical to observed values. With a SSP we can calculate the value of arbitrary ratios at infinite signal to noise on a grid of ages and binary fractions. To estimate the error on the real data, we follow \citet{rosslowe15}, and assume the error on each number count measurement $N$ is $\sqrt{N}$. Thus, for subtypes X and Y with observed number counts $X$ and $Y$, the error\footnote{It is important to note that is a very naive assumption; $X$ and $Y$ are discrete Poisson variables, and their ratio $X/Y$ is not normally distributed. Thus assigning an equal-tailed confidence interval to $X/Y$ via $\sigma_{X/Y}$ is incorrect. As we only wish to compare the rough scales of the spread in model predictions and the typical uncertainty of measured ratios, using Bayesian inference or other methods to construct more accurate confidence intervals is beyond the scope of this paper. However, authors wishing to make quantitive inferences absolutely should make robust error estimates.} of the measured ratio $X/Y$ is
\begin{equation}
\sigma_{X/Y} = \frac{X}{Y}\sqrt{\frac{1}{X} + \frac{1}{Y}}
\end{equation}

\begin{figure*}[ht!]
\centering 
\leavevmode 
\includegraphics[width={0.329\linewidth}]{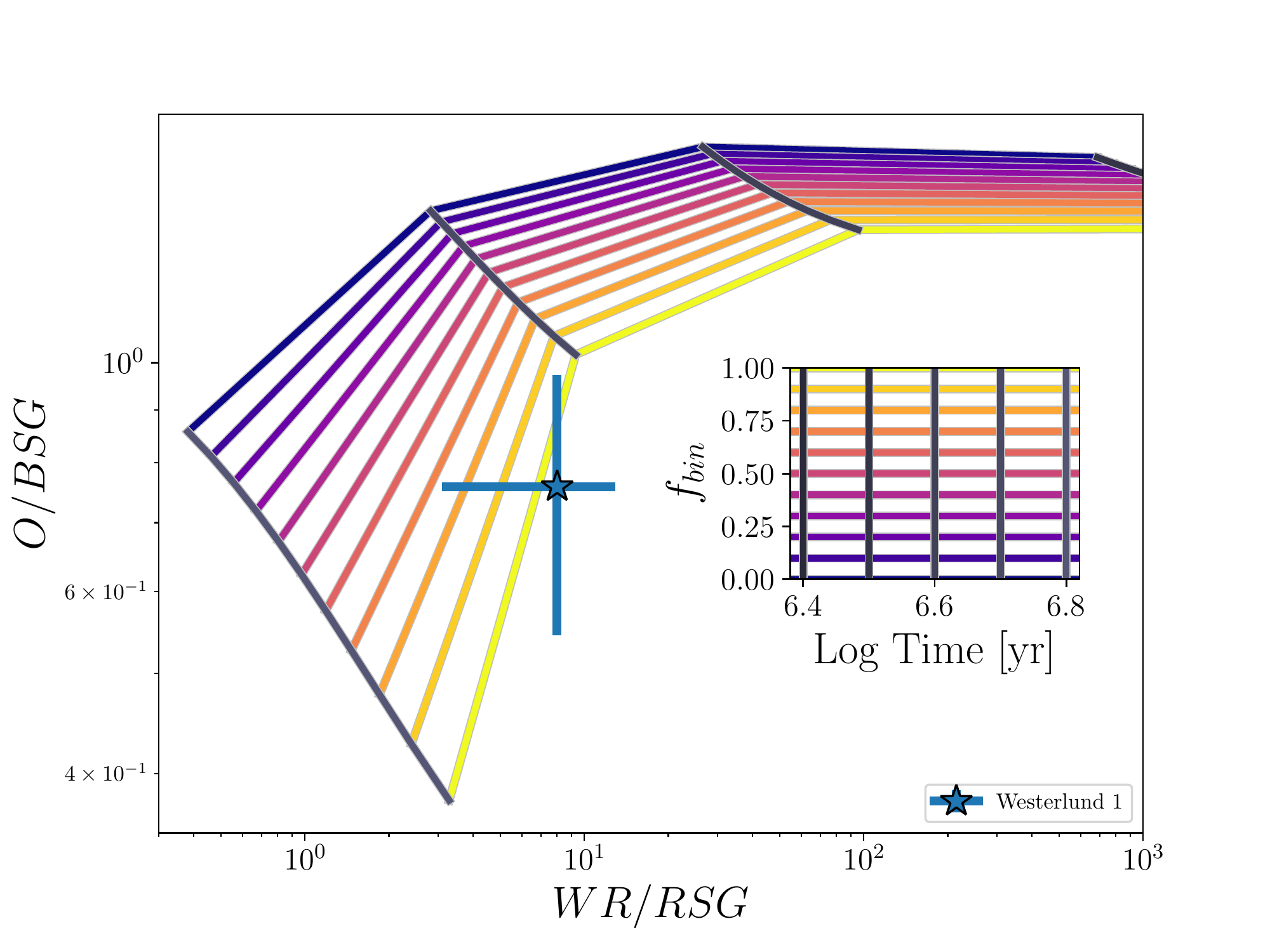} 
\hfil 
\includegraphics[width={0.329\linewidth}]{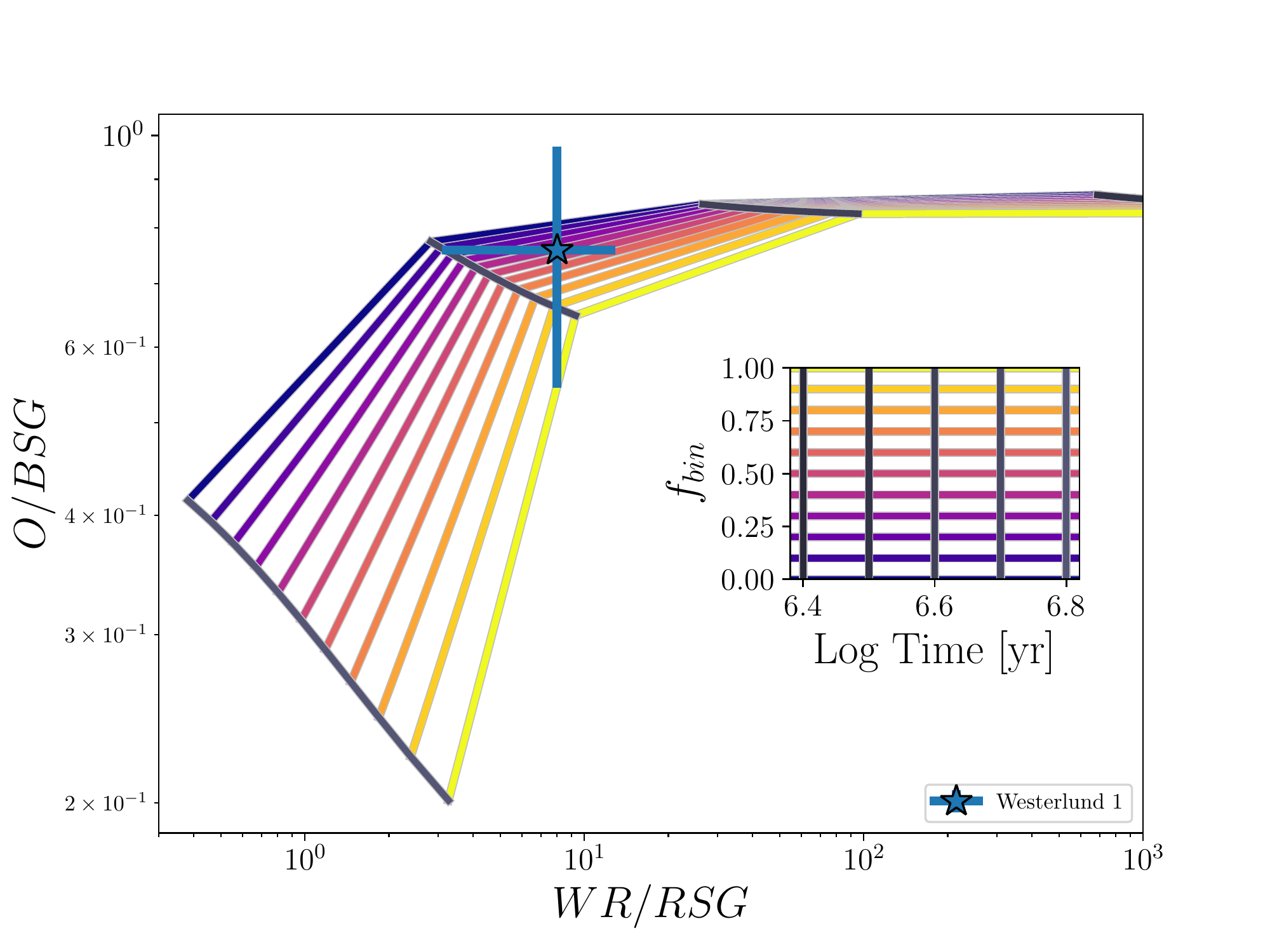}
\hfil
\includegraphics[width={0.329\linewidth}]{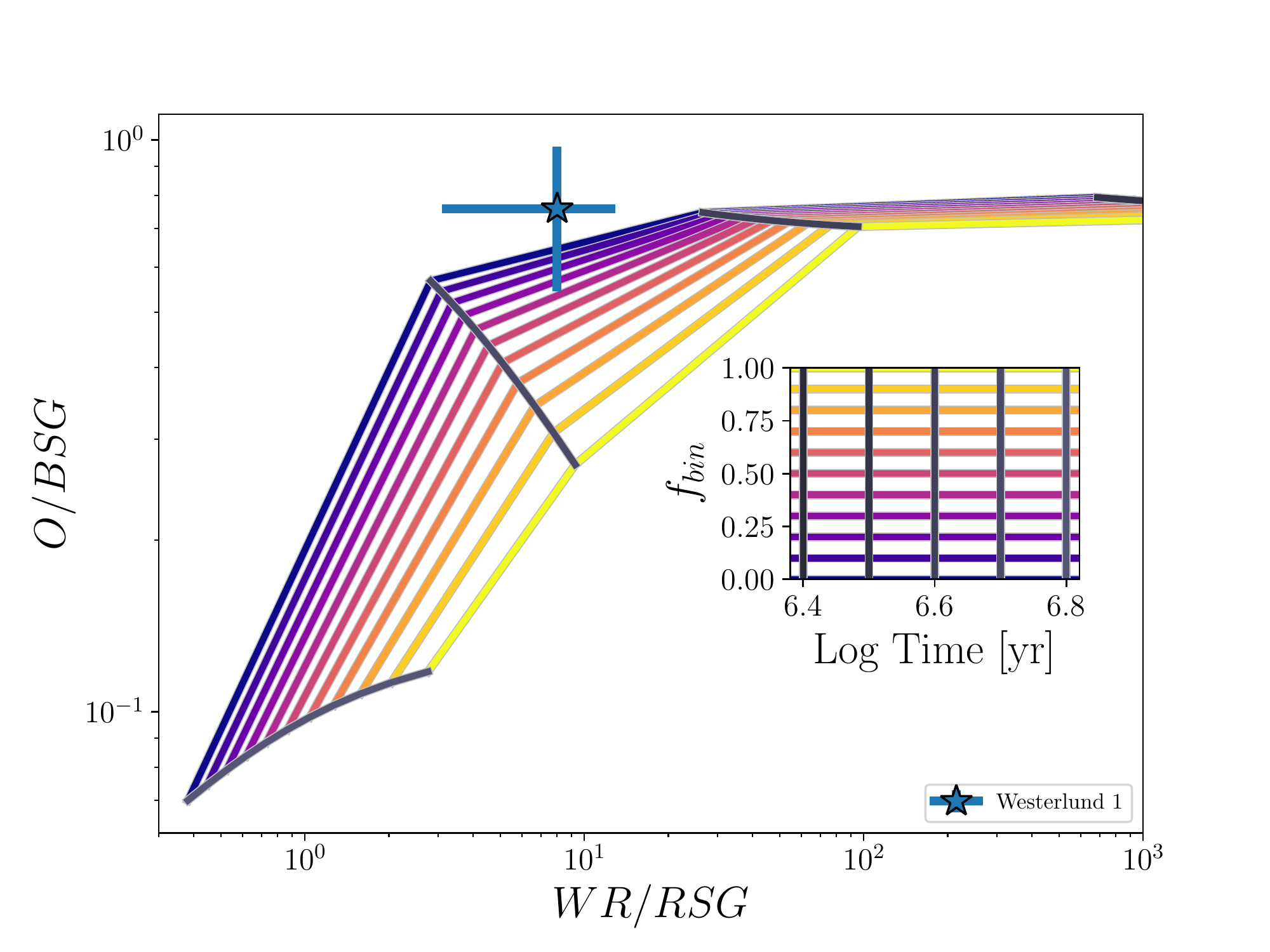}
\caption{{\it Left}: Diagnostic two-ratio plot, applied to the young super star cluster Wd1, which has a cohort of WR, RSG, BSG, and O stars. The inset grid is for reference when interpreting the figure, indicating $f_{bin}$ increases from purple to yellow, and (log) time increases from dark to light grey. {\it Center}: Identical, but assuming completeness limits consistent with the data: WR stars are complete down to $\log(L)=5.1$, all O stars are supergiants ($\log(L)\geq4.9$), and the RSG sample is complete ($\log(L)\geq4.9$). {\it Right}: Identical, but assuming an overly conservative completeness limit of $\log(L)=5.2$ for all species.} \label{fig:two_ratio_ssp}
\end{figure*}

The following examples illustrate a few possibilities. We invite the reader to make use of our publicly available code to explore the entire parameter space to develop diagnostic plots suitable to their dataset. We note that, while the assumption that real star clusters are true SSPs is suspect \citep{gossage18}, there are a variety of open clusters with an age spread less than the time resolution of our SSPs.

We first consider a young ($<10$ Myr) solar metallicity population, which has a wealth of both main sequence and evolved massive stars. Many Galactic super star clusters are this age, and are massive enough to have well-populated IMFs out to tens of $M_\odot$. Westerlund 1 (Wd1) in particular is notable for being a $M_*\approx5\times10^4$ $M_\odot$ cluster \citep{andersen17}, with a well studied cohort of evolved massive stars \citep{clark05,crowther06}. Notably, \citet{crowther06} used the diagnostic ratio $WR/(RSG+YHG)$ (where $YHG\equiv$yellow hypergiant) to estimate the age of Wd1 as ${\sim}4.5-5$ Myr. We can now directly compare the BPASS models with the observed number count data to determine Wd1's age while allowing for a variable $f_{bin}$.

To obtain constraints on both quanitities, we need to use two star count ratios. When the grids of $f_{bin}$ and age are projected into the ratio space, the ensuing topology can be complicated, making inferrence difficult. Thus it is critical to choose ratios such that the grid of parameters remains somewhat orthogonal (or at the very least, non-degenerate). The left panel of Figure \ref{fig:two_ratio_ssp} shows the predicted values for $O/BSG$ vs. $WR/RSG$ at solar metallicity for $\log{t}$ between 6.4 and 6.8 (2.5 and 6.3 Myr). The inset shows the grid of parameter values, with $f_{bin}$ increasing from purple to yellow, and time increasing from dark to light grey. At the earliest times, the model grid is highly degenerate. However, in the latest time bins, different values of age and $f_{bin}$ yield a large spread of possible ratio values. Using this diagram, one can plot the observed values of the two ratios, find the colors of lines that the point intersects, and use the inset to directly determine a binary fraction and age. 

The data from Wd1 are indicated with the blue cross, which assumes $\sqrt{N}$ errors. From \citet{clark05} and \citet{crowther06}, we count 22 O stars, 29 BSGs, 24 WR stars, and 3 RSGs. We assume here that the sample in \citet{clark05} and \citet{crowther06} is complete down to $\log(L)=4.9$ for all subtypes but O stars, and that all O stars have been found.  Note that the O star sample as reported consists mostly of supergiants. At the approximate age of Wd1, the main sequence turnoff is $\sim$30-40 $M_\odot$, implying there are still undetected main sequence O stars; we discuss the implications of an incomplete O star sample in the next section. It may be the case that some WRs and RSGs are obscured by dust, and thus that the sample of evolved stars is also incomplete. However, radio studies of Wd1 have failed to produce previously unknown dust-enshrouded members \citep[e.g., ][]{andrews18,dougherty07}. 

With no further assumptions made about the completeness of the data, we can infer that Wd1 has a high binary fraction of $f_{bin}\gtrsim0.7$, and an age of 5-6.3 Myr. The age is consistent with previous studies, while the measured binary fraction is consistent with the results from \citet{sana12}. This example highlights the importance of constructing ratios that are appropriate for the population under consideration. For a younger (${\sim}3$ Myr) cluster, the errors on the measurements of $WR/RSG$ and $O/BSG$ would have made any inferrence impossible. However, the grid covers a much larger area in ratio space at the latest time shown, implying $WR/RSG$ vs. $O/BSG$ is an even more sensitive metric at older ages.

\begin{figure}[!ht]
\plotone{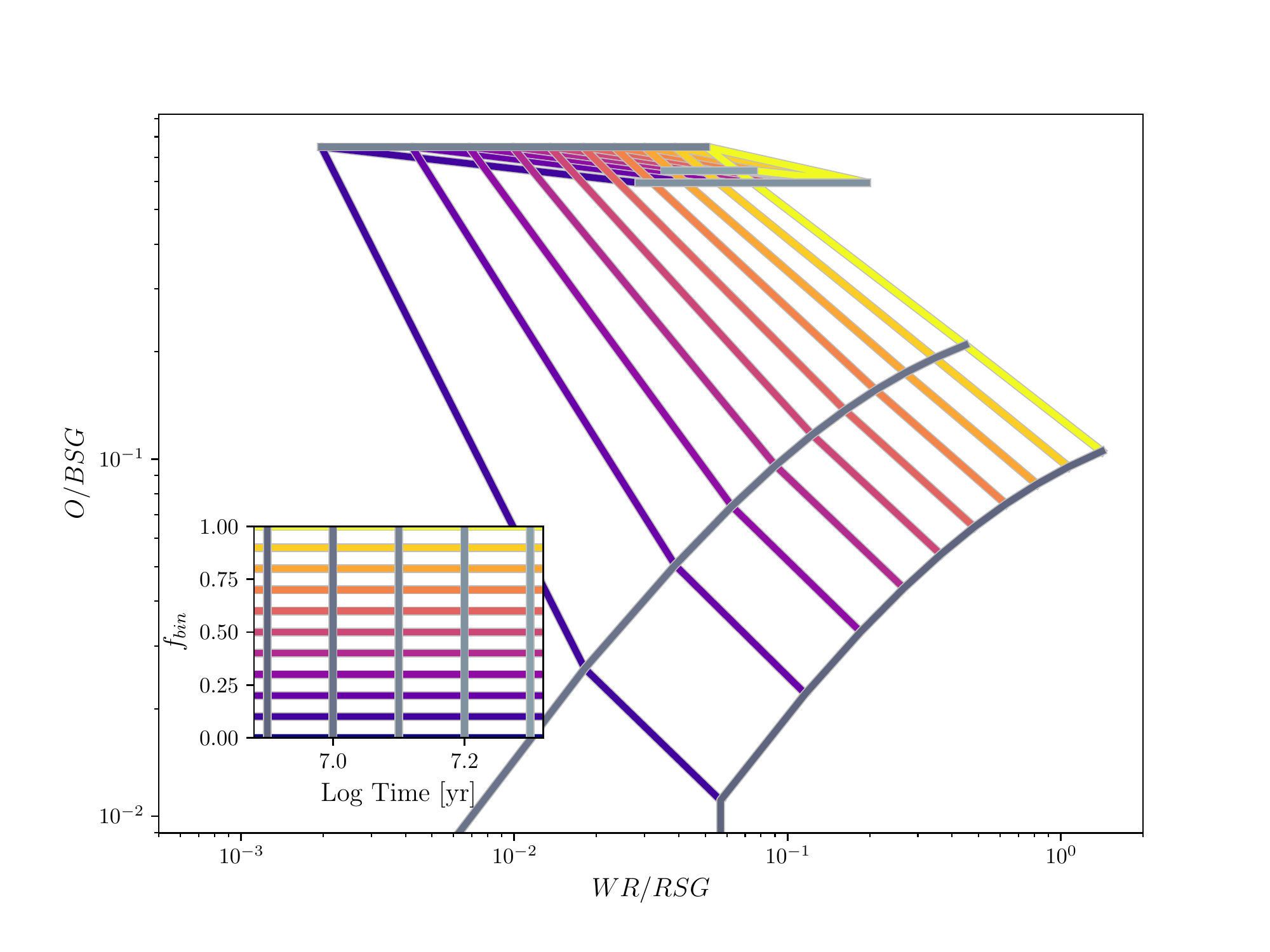}
\caption{Similar diagnostic plot to Figure \ref{fig:two_ratio_ssp} for a theoretical $\sim 20$ Myr cluster. Note: at this age, no single-star WRs are left, so the $f_{bin}=0$ portion of the model isn't visible.}\label{fig:two_ratio_oldssp}
\end{figure}

\begin{figure*}[!ht]
\plotone{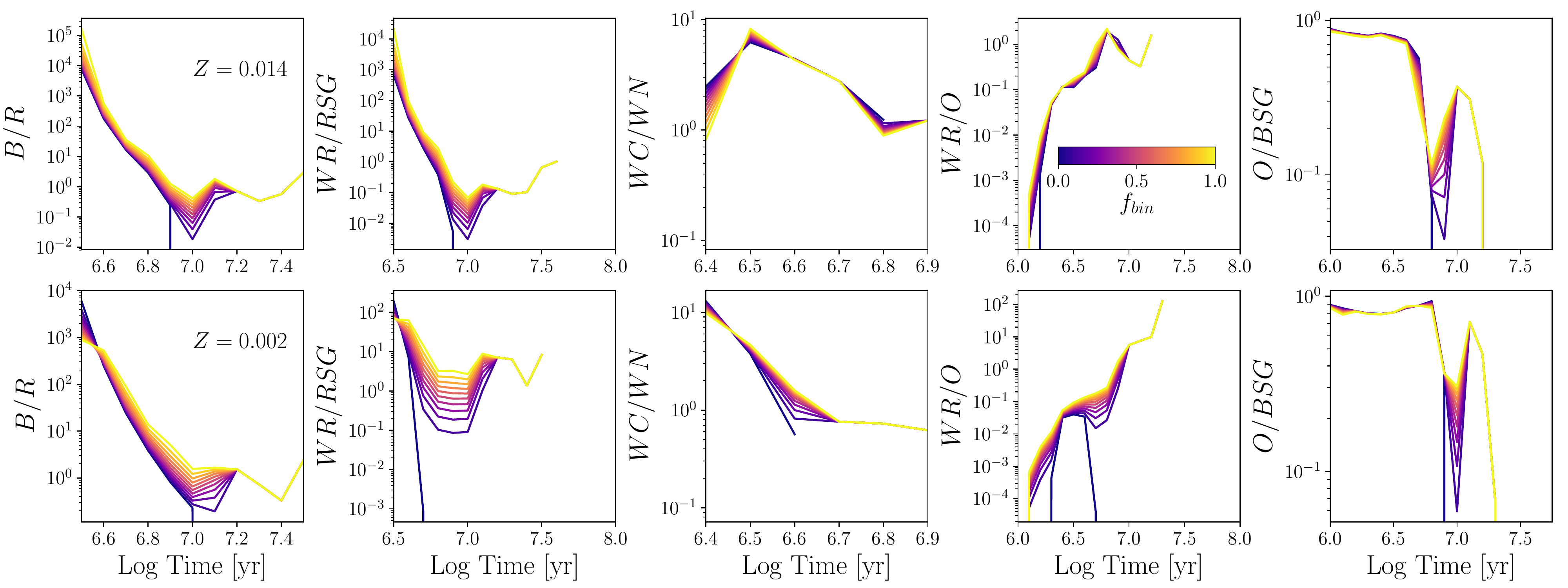}
\caption{Identical to Figure \ref{fig:allratios_allmet}, except assuming a minimum luminosity of $\log(L)=5.2$ for all subtypes.}\label{fig:allratios_incomplete}
\end{figure*}

We next consider the same ratios in a cluster with an age of approximately 10-20 Myr ($\log{t}\approx7-7.3$) in Figure \ref{fig:two_ratio_oldssp}. We calculate $WR/RSG$ and $O/BSG$ for $\log{t}$ between 6.9 and 7.4 (8 and 25 Myr), and binary fractions between 0 and 1. A reference grid to aid in the interpretation of the plot is shown in the inset. Once again these ratios separate well in this space for all but the latest times shown, and span multiple orders of magnitude, implying that this is an incredibly useful diagnostic plot. Unfortunately, all of the WR stars that contribute to $WR/RSG$ at this age are the heretofore mostly undiscovered products of binary evolution. As discussed previously, stripped ``old'' WR stars that evolve from lower-mass progenitors are incredibly rare and most may not be as spectroscopically obvious as their more massive single-star cousins. In a spectroscopic survey of $h+\chi$ Persei, a 13-14 Myr double cluster, \citet{slesnick02} reached the main sequence down to a spectral type of A1, and found no low-mass WR candidates; however, not all of the bright, blue stars were observed. Future deep observing campaigns and theoretical work may yet reveal these stars; unfortunately, this plot remains unusable until then.

While these are only two examples using the same set of ratios, any combination of stellar types (and associated completeness limits, which we discuss further in \S\ref{sec:luminosity}), metallicites (within the set of metallicities modelled with BPASS), and $f_{bin}$ can be used. We make all code that we wrote to generate these plots available online and provide additional examples at \href{https://github.com/tzdwi/Diagnostics}{\tt https://github.com/tzdwi/Diagnostics}.

\section{Accounting for Observational Completeness in Real Samples}\label{sec:luminosity}

Unfortunately, real surveys of Galactic and extragalactic populations are hindered by issues such as source confusion/crowding, inconsistent source classification, and incompleteness, the last of which we discuss here. Throughout this work, we have assumed that WR stars and supergiants were limited to $\log(L)\geq4.9$, while stars on the main sequence could be found and classified with infinite precision. 

Correcting for incompleteness in observed samples of massive stars is a difficult task that can introduce additional uncertainty into an already highly uncertain measurement. We can instead account for incompleteness in our synthetic populations by increasing the lower luminosity criteria for individual subtypes. Consider, for example, a spectroscopic survey that is only complete down to $\log(L)=5.2$, below which no stars are classified. Figure \ref{fig:allratios_incomplete} shows identical ratios to Figure \ref{fig:allratios_allmet} after accounting for this completeness limit. Compared to Figure \ref{fig:allratios_allmet}, almost all of the ratios become far less dependent on $f_{bin}$, especially at later times. This means that results obtained using ratios applied to incomplete samples will be less affected by binaries that weren't accounted for in the analysis. Conversely, incompleteness makes the task of simultaneously measuring age and $f_{bin}$ much more difficult. 

In practical terms, survey completeness is often expressed as a limiting magnitude in an optical bandpass, which can be transformed into a limiting absolute magnitude in that band. Thanks to blackbody physics, a magnitude limit like this corresponds to different bolometric luminosity limits for different spectral types. Therefore the above example is a toy model. In actuality, a survey will be more sensitive to bluer stars, and thus the exact luminosity limits should be carefully chosen to match the limits of the data.

As an example, we return again to Wd1. The catalog of \citet{clark05} mostly does not include main sequence O stars; instead, most of the OBA stars they found belong to the cluster's supergiant cohort ($\log(L)\geq4.9$). Using a combination of narrow- and broad-band imaging and spectroscopy to hunt for WR stars, \citet{crowther06} only find stars brighter than $M_{bol}=-8.2$ ($\log(L)\approx5.18$). In the center panel of Figure \ref{fig:two_ratio_ssp}, we again plot $O/BSG$ vs. $WR/RSG$ for our theoretical populations and for Wd1. However, in this figure, we impose a minimum WR luminosity of 5.1, and a minimum O luminosity of 4.9. These changes yield a model grid that predicts smaller values for both $O/BSG$ and $WR/RSG$. The implied age of Wd1 is now slightly younger at 4-5 Myr, while the uncertainty in the observed ratios makes it impossible to measure a value of $f_{bin}$. We note that \citet{crowther06} made no such correction for completeness when using $WR/(RSG+YHG)$ to estimate the age of Wd1. 

The right panel of Figure \ref{fig:two_ratio_ssp} is identical to the left and center panels, but assumes an overly conservative completeness limit of $\log(L)=5.2$ for all species. The model grid predicts smaller values for both ratios, an age consistent with \citet{crowther06}, and an upper limit for the binary fraction of $f_{bin}\lesssim0.4$, which is inconsistent with current measurements of $f_{bin}$. This stresses the importance of choosing completeness limits that are consistent with the data, rather than relying on conservative assumptions.

Ultimately, completeness limits combined with uncertain values of $f_{bin}$ can affect the theoretical values of star count ratios, hindering measurements of age or $f_{bin}$. Thus, great care must be taken to ensure that the Poisson noise of the measurement is smaller than the anticipated effect of binaries or incompleteness. For example, if precision of $\sim0.1$ dex in a star count ratio $R=X/Y$ is required to measure a value of $f_{bin}$ to within 0.1, and the expected value of the $R$ is $R\approx1$, the observer should find approximately 100 of each type of star in order to obtain the necessary precision, assuming they use $\sqrt{N}$ errors. Of course, the exact number of stars required changes with the age and metallicity of the population, but, as a rough rule of thumb, sample sizes of $\sim100$s are necessary to make precision measurements. While only the most massive of galactic star clusters have the requisite number of stars, entire galaxies do have enough massive stars. We now turn our attention to calculating star count ratios in galaxies with complex star formation histories. 

\section{Ratios in Complex Star Formation Histories}\label{sec:complex}

\begin{figure}[!ht]
\plotone{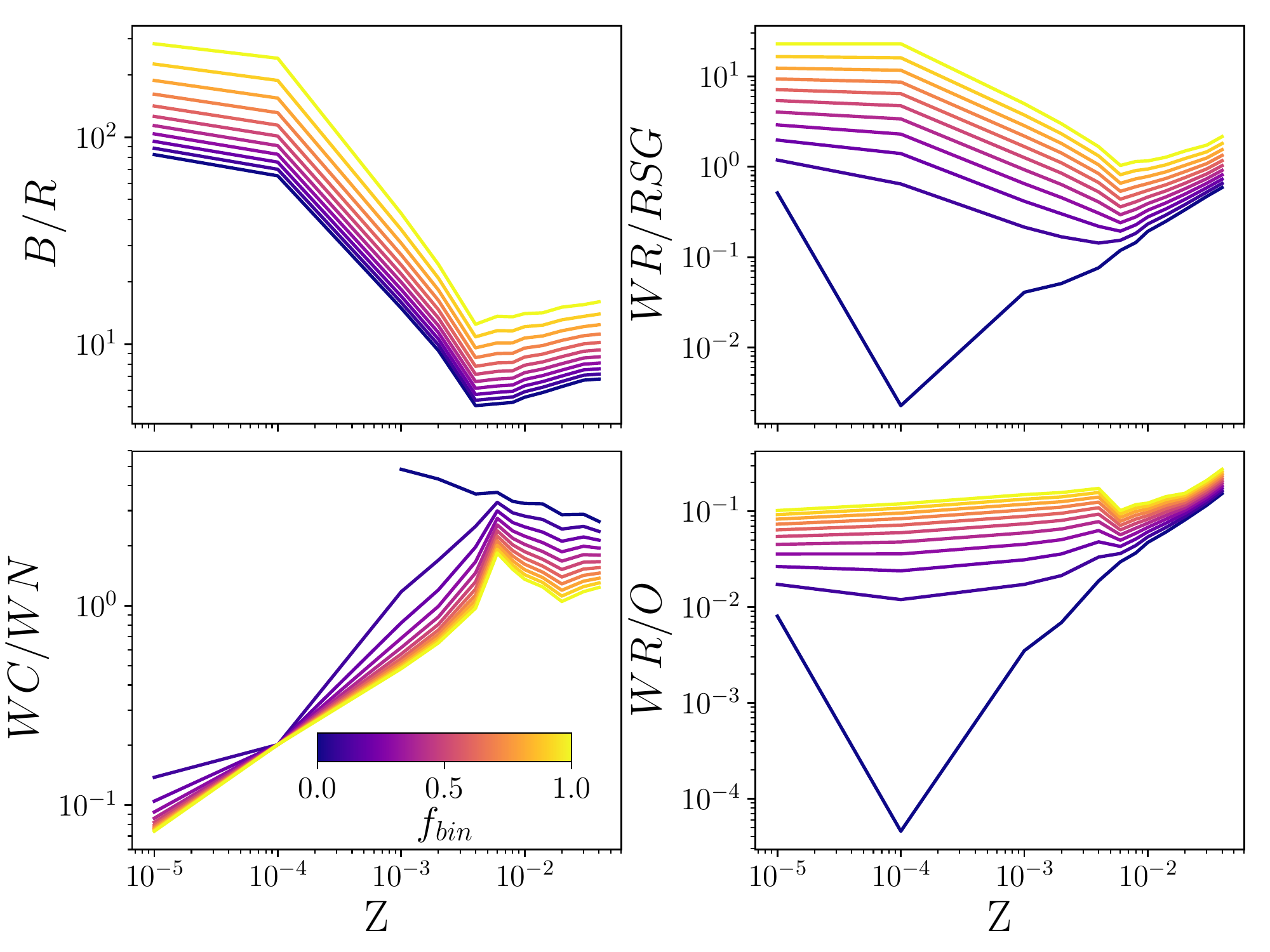}
\caption{$B/R$ (top left), $WR/RSG$ (top right), $WC/WN$ (bottom left), and $WR/O$ for galaxies with constant star formation, and values of $f_{bin}$ between 0 and 1; the color mapping for $f_{bin}$ is identical to previous plots. A minimum luminosity of $\log(L) = 4.9$ is assumed for all subtypes.}\label{fig:const_SFH_ratios}
\end{figure}

While our discussion thusfar has been limited to SSPs, star count ratios are most frequently applied to entire galaxies, where they are used both as a benchmark of the physics implemented in a stellar evolution code and a test of our understanding of these physics. However, galaxies have complex star formation histories (SFHs); the galaxy as we see it today can be seen as the integrated set of countless populations formed between the onset of star formation and today, weighted by the star formation history, $\Psi$. We implement complex star formation histories in our code as follows. The number of a subtype S at time $t$ for a SSP is $S(t)$. The total number seen in a population with star formation history $\Psi(t)$ is
\begin{equation}
S_{tot}(t_{i_{max}}) = \sum_{i=0}^{i_{max}}\Psi(t_i)S(t_i)\Delta t_i
\end{equation}
where $t_i$ is the $i^{\rm th}$ BPASS time bin with width $\Delta t_i$, and $i_{max}$ corresponds to the total age of the population. Because BPASS uses 51 logarithmically spaced time bins from 1 Myr to 100 Gyr, $\Delta t_i$ is thus
\begin{equation}
\Delta t_i = {\begin{cases} 
    10^{6.05} & i = 0 \\
    10^{6.15 + 0.1i} - 10^{6.05 + 0.1i} & 1\leq i\leq51
    \end{cases}}
\end{equation}
We note that the definition of $\Psi$ is such that $\Psi(t_0)$ is the star formation rate 1 Myr {\it ago} (i.e., the youngest BPASS time bin), not the star formation rate when the population is 1 Myr old. In the following, we adopt a constant star formation rate of $\Psi(t) = 1$ $M_\odot$ yr$^{-1}$ following \citet{eldridge17}; however, our code allows for arbitrary SFHs, discretized to the default 51 age bins in BPASS. Note that, though we compute populations assuming constant SFR for all BPASS age bins, populations of massive stars are only sensitive to (at most) the previous 50-100 Myr of star formation, after which point the numbers of massive stars reach an equilibrium.\footnote{Indeed, the ionizing spectra of young populations in starburst galaxies reaches an equilibrium far earlier, at $\sim5$ Myr \citep{kewley01}.} Because of this fact, the implicit assumption of constant metallicity for the entire population, while still unrealistic, is slightly more tenable. 

Figure \ref{fig:const_SFH_ratios} shows $WR/O$, $B/R$, $WR/RSG$, and $WC/WN$ as a function of metallicity in galaxies with constant SFR after allowing the massive star populations to reach equilibrium. We also implement a minimum luminosity of $\log(L)=4.9$ through the remainder of this section, for consistency with extragalactic samples which are typically incomplete below this luminosity. For all four ratios plotted, introducing binary stars adds more than an order of magnitude spread in the predicted values for these ratios at most metallicities. This implies that inferences of metallicity, or star formation rate from number count ratios assuming only single star models are incorrect. However, the order-of-magnitude effect of binaries along with the improvement in statistics afforded by studying galaxies instead of star clusters implies that we can analyze extragalactic populations within the self-consistent framework of BPASS to make qualitative statements about binary populations. 

\begin{figure*}[!ht]
\plottwo{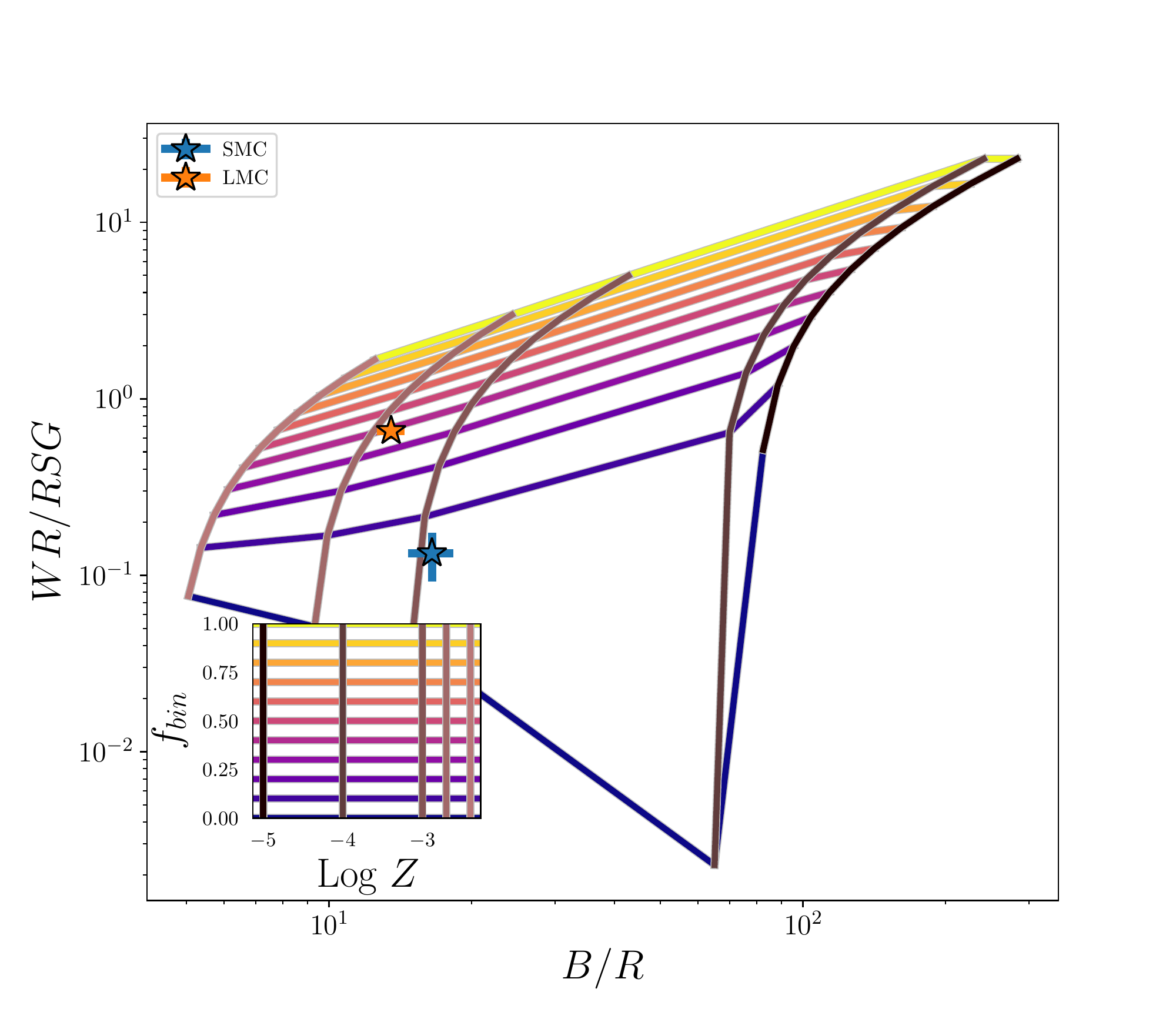}{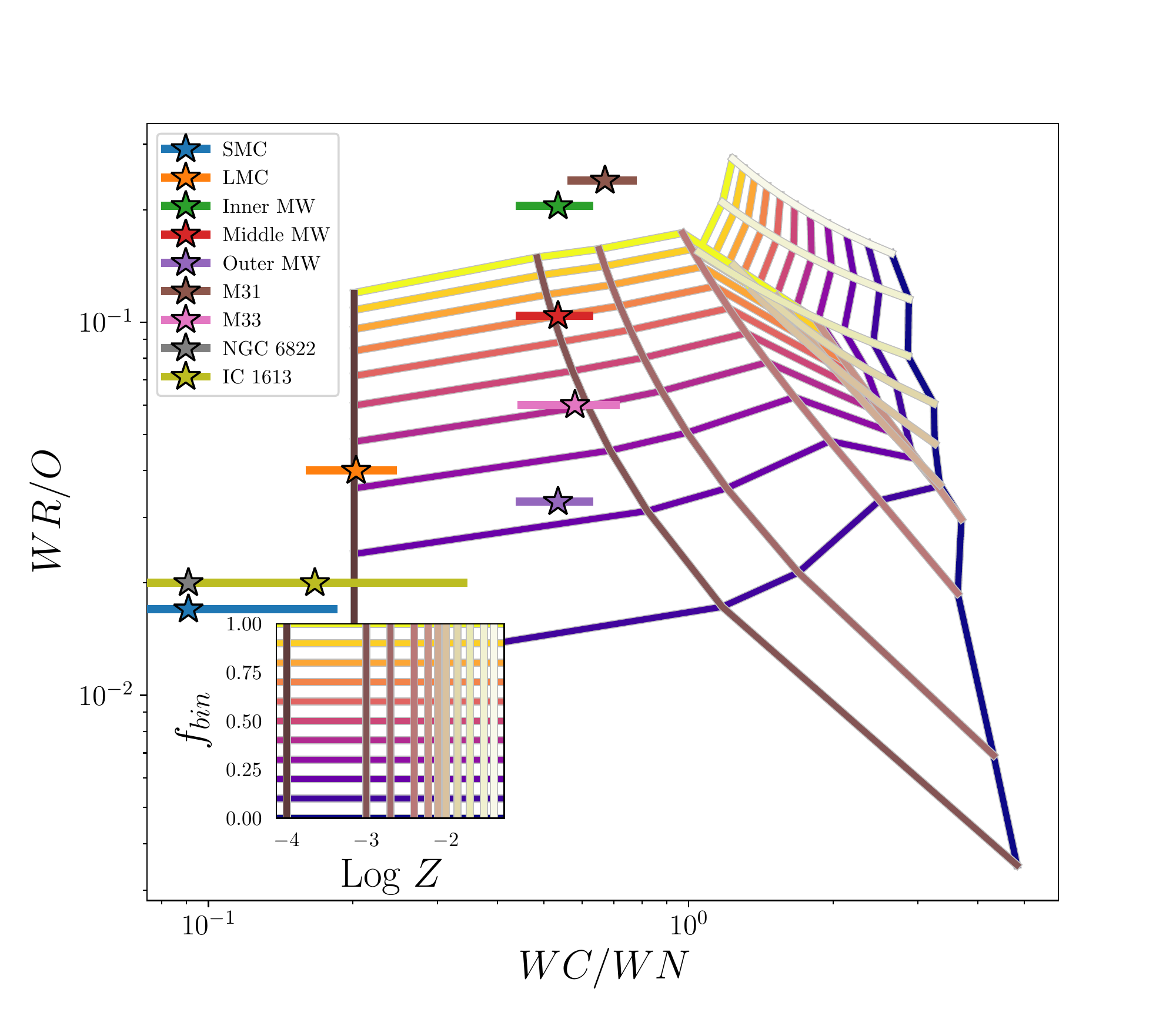}
\caption{$B/R$ vs. $WR/RSG$ (left) and $WR/O$ vs. $WC/WN$ (right) for constant star formation populations with varying metallicity and $f_{bin}$ (metallicity in the right plot is limited to $10^{-5}\leq Z\leq0.004$). The insets are similar references grids to the plots in Figure \ref{fig:two_ratio_ssp}. Observed values of these ratios (including errors where provided by the original authors) for various populations around the local group are plotted --- note that $WR/O$ is equal in NGC 6822 and IC 1613). $B/R$ data are from \citet{massey03}; $WR/RSG$ data are from \citet{massey03} for RSGs, \citet{neugent11} for SMC WRs, and \citet{neugent18} for LMC WRs; $WR/O$ data are from \citet{maeder94}; and $WC/WN$ data are from \citet{rosslowe15} for the Galactic WRs; \citet{armandroff85} in NGC 6822 and IC 1613; and \citet{neugent11} and \citet{neugent12,neugent18} in the Magellanic Clouds, M31, and M33. All subtypes are assumed to have a minimum luminosity $\log(L)=4.9$, including O stars, to account for incompleteness in extragalactic studies.}\label{fig:two_ratio_constSFH}
\end{figure*}

Figure \ref{fig:two_ratio_constSFH} shows $B/R$ vs. $WR/RSG$ and $WR/O$ vs. $WC/WN$, with lines of constant $f_{bin}$ and $Z$ for populations with constant star formation, along with inset reference grids in a similar fashion to Figure \ref{fig:two_ratio_ssp}. We also plot real data from the SMC (blue, $Z=0.002$); LMC (orange, $Z=0.006$); solar neighborhood stars with galactocentric distances of $6\leq R_{GC}<7.5$ kpc (green, supersolar metallicity),  $7.5\leq R_{GC}<9$ kpc (red, $Z\approx Z_\odot$), and $9\leq R_{GC}<11$ kpc (purple, subsolar metallicity); M31 (brown, $Z\approx2Z_\odot$); the inner region of M33 (pink, $Z\approx Z_\odot$); NGC 6822 (yellow, $Z=0.005$); and IC 1613 (grey, $Z=0.002$). $B/R$ data are from \citet{massey03} and $WR/RSG$ data are compiled from \citet{massey03}, \citet{neugent11}, and \citet{neugent12}. $WR/O$ data are all from \citet{maeder94}, who do not report raw numbers to allow us to estimate an error. $WC/WN$ data are from \citet{rosslowe15} in the Milky Way (where we only use their data from the solar circle to compare with the smaller annuli from \citealt{maeder94}); \citet{armandroff85} in NGC 6822 and IC 1613; and \citet{neugent11} and \citet{neugent12,neugent18} in the Magellanic Clouds, M31, and M33. \citet{neugent12} report both $\sqrt{N}$ and rigorous asymmetric errors based on their completeness limits. For consistency between varying sources of data we only use the $\sqrt{N}$ errors, but note that quantitative analyses should adopt rigorous error calculations. We tabulate the raw numbers used to calculate the ratios and errors, as well as the value of $WR/O$ from \citet{maeder94} in Table \ref{tab:raw}.

\begin{deluxetable}{lcccccc}
\tabletypesize{\normalsize}
\tablecaption{Raw counts of massive stars used to calculate ratios and errors, as well as the value of $WR/O$ from \citet{maeder94}\label{tab:raw}.}
\tablehead{\colhead{Galaxy} &
\colhead{$BSG$} & \colhead{$RSG$} & \colhead{$WR$} & \colhead{$WC$} & \colhead{$WN$} & \colhead{$WR/O$}} 
\startdata
SMC       & 1484 & 90  & 12  & 1   & 11  & 0.017  \\
LMC       & 3164 & 234 & 154 & 26  & 128 & 0.04   \\
Inner MW  & ---  & --- & --- & --- & --- & 0.205  \\
Middle MW & ---  & --- & --- & 46\tablenotemark{a}  & 86\tablenotemark{a}  & 0.104  \\
Outer MW  & ---  & --- & --- & --- & --- & 0.033  \\
M31       & ---  & --- & --- & 62  & 92  & 0.24   \\
M33       & ---  & --- & --- & 26  & 45  & 0.06   \\
NGC 6822  & ---  & --- & --- & 1   & 11  & 0.02   \\
IC 1613   & ---  & --- & --- & 1   & 6   & 0.02   \\
\enddata
\tablecomments{\scriptsize \tablenotetext{a}{While \citet{rosslowe15} report $WC/WN$ for the ``Inner'' and ``Outer'' Milky Way, we only use $WC/WN$ data from the ``Middle'' to plot against $WR/O$, as annuli in galactocentric distance used by \citet{maeder94} are much thinner, and don't overlap with the inner and outer annuli from \citet{rosslowe15}}} 
\end{deluxetable}

The first thing that we see is that $f_{bin}$ and $Z$ projected into the ratio spaces are well separated across multiple orders of magnitude, and are roughly orthogonal; i.e., $WR/RSG$ and $WR/O$ are good tracers of $f_{bin}$, while $B/R$ and $WC/WN$ trace metallicity. Thus, changes in the physics in BPASS that only affect the lifetime of one evolutionary phase (e.g., an implementation of meridional circulation, which would increase the BSG lifetime, \citealt{eldridge17}), will predominantly change the absolute inferred $f_{bin}$ of $Z$; the relative values inferred are still valid.

When compared to data, we see that the metallicity that one might infer based on these grids is systematically lower than the true metallicity of the galaxies (e.g., on the left-hand plot, one might assume that $Z_{SMC} \approx 0.001$ and $Z_{LMC}\approx0.002$; the right-hand plot shows an even worse correspondence). This may be due to at least one of a few possibilities:
\begin{enumerate}
    \item The data are subject to inconsistent classification and difficult-to-quantify completeness-limits.
    \item The completeness of these samples extends to lower luminosities than are assumed in the model grids. For example, a lower luminosity cutoff in the models would add more WNs than WCs (because lower luminosity WRs will be lower mass, and thus more likely to be WNs), and more Os than WRs (due to the relative lifetimes and abundances of both subtypes), shifting the grid down and to the left in the right hand plot.
    \item The mass loss through both single- and binary- star channels predicted by BPASS is incorrect. Mass loss will govern the age at which massive stars transition between various evolutionary stages, and $WC/WN$ and $B/R$ are both sensitive to these lifetimes. 
    \item The effects of rotation included in BPASS are incorrect. BPASS only uses approximate physics to simulate rotation. Including e.g., meridional circulation, would increase the number of BSGs, which would then increase the metallicity inferred by $B/R$. Future work will examine the effects of rotation using stellar evolution codes that adopt more detailed treatments of rotational phenomena.
\end{enumerate}

\begin{figure}[!ht]
\plotone{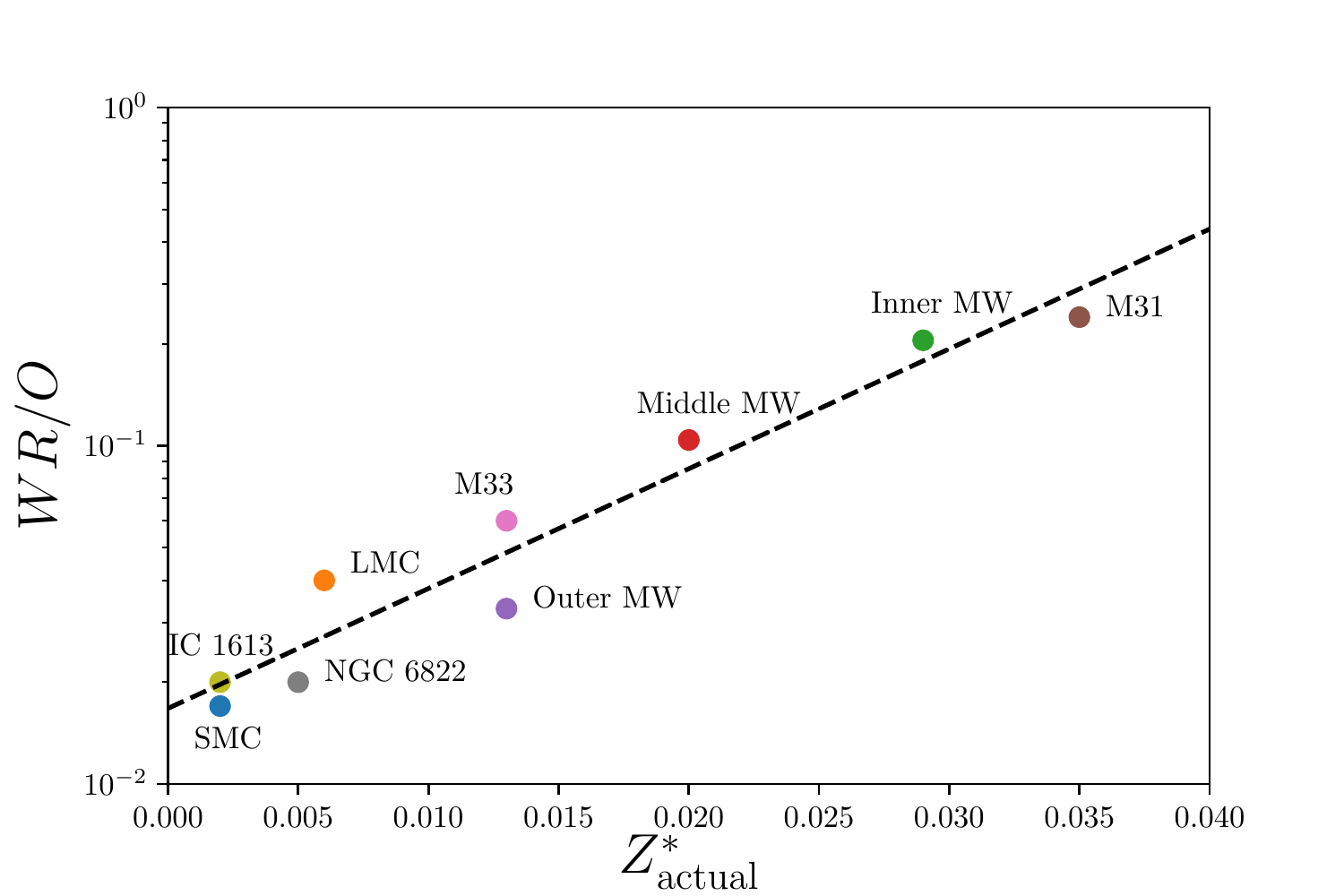}
\caption{$WR/O$ vs. actual metallicity for galaxies in the local group. All data are from \citet{maeder94}; those authors assumed $Z_\odot=0.02$. For self-consistency, we only report $Z$ as listed in Table 6 of \citet{maeder94}; thus we put an asterisk in the x label. The dashed black line is a linear fit to $\log{(WR/O)}$ vs. $Z^*_{\rm actual}$.}\label{fig:wro_z}
\end{figure}

Assuming that, while the absolute values of metallicity or $f_{bin}$ or $Z$ inferred from these plots will change depending on the exact physical prescriptions, the relative values will remain largely consistent, we see that $f_{bin}$ in all of these galaxies appears to increase with actual metallicity. If $WR/O$ is indeed a good tracer of $f_{bin}$, this should also be apparent when plotting $WR/O$ vs. $Z$. Figure \ref{fig:wro_z} shows the values of $WR/O$ and $Z$ as listed in Table 6 of \citet{maeder94}, along with a linear fit to $\log{(WR/O)}$ vs. $Z_{\rm actual}$. Note that we use an asterisk in the x-axis label because the metallicities reported in \citet{maeder94} assume $Z_\odot=0.02$, instead of 0.014 \citep{asplund09}. The fit clearly shows the trend that $WR/O$, and thus $f_{bin}$, increases with $Z$ (under the assumptions above).

While more work is necessary both in the BPASS models and in observational studies in order to quanitify the exact relationship between $f_{bin}$ and $Z$, this is still a very intriguing finding. A correlation between $f_{bin}$ and $Z$ for massive stars has not previously been reported. \citet{raghavan10} considered low-mass stars, and found that metal-poor main sequence stars with $0.625\leq B-V\leq1.0$ were more likely to have stellar companions at a confidence of 2.4$\sigma$. If the binary fraction instead increases with metallicity for more massive stars, this would point to a fundamental difference in how stellar binaries are formed in different mass regimes. Indeed, this result makes intuitive sense; increased metal line cooling in a molecular cloud would make the cloud cooler and denser, while carrying away none of the angular momentum and making the formation of binaries more likely. We note that this argument applies only to the formation of binary systems, and not their {\it evolution}. Indeed, at lower metallicity, binaries are expected to be {\it found} at closer orbital separations due to weaker stellar winds being less efficient at losing angular momentum, thus increasing orbital separations \citep{demink08,linden10}. BPASS does include these physics, but this result is merely a statement about the natal, intrinsic value of $f_{bin}$.

It is important to note  that there are caveats to this result: if the binary fraction is dependent on metallicity, then it is also likely that the period and mass ratio distributions are as well, while the BPASS inputs implicitly assume only the binary parameter distributions from \citet{moe17}. Future work will include varying the period and mass ratio distributions and considering any potential dependence on metallicity. We also make no attempt at quantifying this putative relationship between $f_{bin}$ and $Z$, as such a result is not the focus of this work, and will require dedicated observational and theoretical study.

\section{Summary and Conclusion}\label{sec:conclusion}

We summarize our results as follows:
\begin{itemize}
    \item When applied to SSPs, accounting for binary effects when using star count ratios is incredibly important. If the binary fraction isn't known, an order-of-magnitude spread can be introduced into the theoretical prediction.
    \item We find including binaries and imposing a completeness limit can both introduce $\gtrsim0.1$ dex changes in the inferred $\log({\rm age/Myr})$ of star clusters.
    \item Similar incompleteness and binary effects can manifest themselves in more complex systems of stars. However, because star count ratios can be subject to fewer systematics and small-number statistics, proper treatment of stellar binaries can yield interesting results. 
    \item Combinations of star count ratios can be used to indirectly measure the massive star binary fraction, provided the data have well understood completeness limits and assumed errors. This method works in large or distant populations where direct measurement of $f_{bin}$ via spectroscopic studies of individual stars is otherwise impossible. Where direct measurements are also possible, this method can also illuminate where the BPASS stellar evolution physics can be improved.
\end{itemize}

\acknowledgments

This work was supported by NSF grant AST 1714285 to E.L.

The authors thank J. J. Eldridge for their help and advice, in addition to their work on the BPASS project. The authors also thank the anonymous referee for their insightful comments and feedback.

T.D.W. thanks N. N. Sanchez for useful discussions about metallicity effects.

This work made use of v2.2.1 of the Binary Population and Spectral Synthesis (BPASS) models as described in \citet{eldridge17} and \citet{stanway18}.

This work made use of the following software:

\vspace{5mm}

\software{Astropy v2.0.3 \citep{astropy13,astropy18}, h5py v2.7.1, Matplotlib v2.1.2 \citep{Hunter:2007}, makecite \citep{makecite18}, NumPy v1.14.1 \citep{numpy:2011}, Python 3.5.1}

\bibliography{DL_bib}
\bibliographystyle{aasjournal}

\end{document}